\documentclass[12pt,a4paper]{article}

\usepackage[margin=2cm]{geometry}
\usepackage[hidelinks]{hyperref}
\usepackage{graphicx}
\usepackage{amssymb, amsmath, amsfonts, bm, mathrsfs,amsthm}
\usepackage{siunitx}
\usepackage[utf8]{inputenc}
\usepackage{cleveref}
\usepackage{siunitx}
\begin{document}

\setlength{\parindent}{0pt}
\setlength{\parskip}{.5em}

\renewcommand{\Re}[0]{\operatorname{Re}}
\renewcommand{\Pr}[0]{\operatorname{Pr}}
\newcommand{\Fr}[0]{\operatorname{Fr}}
\newcommand{\eps}[0]{\varepsilon}
\newcommand{\ehat}[0]{\hat{e}}
\newcommand{\ohat}[0]{\hat{\omega}}
\newcommand{\nhat}[0]{\widehat{n}}
\newcommand{\that}[0]{\widehat{t}}
\renewcommand{\J}[0]{\mathcal{J}}
\renewcommand{\K}[0]{\mathcal{K}}
\newcommand{\E}[0]{\mathcal{E}}
\newcommand{\U}[0]{\mathcal{U}}
\renewcommand{\O}[0]{\mathcal{O}}
\newcommand{\notvc}[1]{#1}
\newcommand{\br}[1]{\left(#1\right)}
\newcommand{\sbr}[1]{\left[#1\right]}
\newcommand{\at}[1]{\left.#1\right|}
\newcommand{\pd}[0]{\partial}
\newcommand{\sech}[0]{\operatorname{sech}}

\title{On the stability of isothermal shocks in black hole accretion disks}

\author{
Eric W. Hester,$^{1}$\thanks{E-mail: ehester@math.ucla.edu}
\and Geoffrey M. Vasil,$^{2}$
\and Martin Wechselberger,$^{2}$ 
}
\date{
% List of institutions
\footnotesize
$^{1}$Department of Mathematics, The University of California, Los Angeles, \\
520 Portola Plaza 90024, California, United States of America\\
$^{2}$School of Mathematics and Statistics, The University of Sydney,\\
Camperdown 2006, NSW, Australia\\[1em]
\today}
\maketitle

\begin{abstract}
Most black holes possess accretion disks. 
Models of such disks inform observations and constrain the properties of the black holes and their surrounding medium.
Here, we study isothermal shocks in a thin black hole accretion flow.
Modelling infinitesimal molecular viscosity allows the use of multiple-scales matched asymptotic methods. 
We thus derive the first explicit calculations of isothermal shock stability.
We find that the inner shock is always unstable, and the outer shock is always stable.
The growth/decay rates of perturbations depend only on an effective potential and the incoming--outgoing flow difference at the shock location. 
We give a prescription of accretion regimes in terms of angular momentum and black hole radius.
Accounting for angular momentum dissipation implies unstable outer shocks in much of parameter space, even for realistic viscous Reynolds numbers of the order $\approx 10^{20}$.
\end{abstract}

\section{Introduction}
The classical models of black hole accretion disk theory \cite{NovikovAstrophysicsBlackHoles1973,
PringleAccretionDiscsAstrophysics1981,
ChakrabartiTheoryTransonicAstrophysical1990,
PapaloizouTheoryAccretionDisks1995,
AbramowiczFoundationsBlackHole2013} predict transonic flows and steady shocks \cite{FukueTransonicDiskAccretion1987,
ChakrabartiStandingShocksIsothermal1989,
ChakrabartiStandingShocksIsothermal1990,
ChakrabartiStandingRankineHugoniotShocks1989,
AbramowiczStandingShocksAdiabatic1990,
ChakrabartiSmoothedParticleHydrodynamics1993,
CaditzAdiabaticShocksAccretion1998}.
Understanding these shocks allows us to infer large scale properties of the disk and black hole \cite{LeSelfconsistentModelFormation2004,LeParticleAccelerationProduction2005,LeParticleAccelerationAdvectiondominated2007}.
The key remaining question is which, if any, of these shocks are stable?
Previous works analyse shock stability using Rankine-Hugoniot conditions in the inviscid Euler equations
\cite{NakayamaHydrodynamicInstabilityAccretion1992,
NakayamaDynamicalInstabilityStanding1994,
YangShockStudyFully1995,
ChakrabartiGlobalSolutionsViscous1996,
YuanSonicPointsShocks1996,
LeSelfconsistentModelFormation2004,
FukumuraIsothermalShockFormation2004,
GuNonaxisymmetricInstabilitiesShocked2003,
GuNonaxisymmetricInstabilitiesShocked2006,
NagakuraGeneralRelativisticHydrodynamic2008,
LeStandingShockInstability2016}.
While these techniques give important qualitative information, 
previous efforts only produced bounds on the growth rates for shock stability \cite{NakayamaDynamicalInstabilityStanding1994}.

This paper uses singular perturbation theory to give precise formulae for the shock growth rates, and finds that they assume the previous lower bound.
That is, we consider accretion of a viscous fluid in the limit of vanishing viscosity.
While this approach may seem more complicated and less predictive, we show that it is just the opposite.
Multiple scales matched asymptotic analysis shows that inner shocks are unstable and outer shocks are stable (agreeing with 
\cite{NakayamaHydrodynamicInstabilityAccretion1992,
NakayamaDynamicalInstabilityStanding1994,
YangShockStudyFully1995,
ChakrabartiGlobalSolutionsViscous1996,
YuanSonicPointsShocks1996,
LeSelfconsistentModelFormation2004,
FukumuraIsothermalShockFormation2004,
GuNonaxisymmetricInstabilitiesShocked2003,
GuNonaxisymmetricInstabilitiesShocked2006,
NagakuraGeneralRelativisticHydrodynamic2008,
LeStandingShockInstability2016}),
and provides the first explicit formulas for the stability of such shocks.
Furthermore, the next-order perturbations imply tighter constraints on the already narrow parameter regimes of steady shocks.
We support all mathematical arguments with a Mathematica notebook, and validate our calculations using simulations with the Dedalus spectral code \cite{BurnsDedalusFlexibleFramework2020}.
The mathematical and computational techniques of this paper are widely applicable, and promise to improve understanding of shock phenomena in astrophysical problems.

\subsection{Background and summary}
\label{sec:background}
Black hole accretion must be transonic.
Accretion flows can only pass the sound barrier at special sonic points \cite{BondiSphericallySymmetricalAccretion1952} ---
the same sonic points discovered for the reverse process of stellar winds by \cite{ParkerDynamicsInterplanetaryGas1958}.
Following the first general relativistic disk models \cite{NovikovAstrophysicsBlackHoles1973,
PageDiskAccretionBlackHole1974,
AbramowiczRelativisticAccretingDisks1978},
simpler models predicted multiple sonic points in black hole accretion disks
\cite{AbramowiczRotationinducedBistabilityTransonic1981, 
LoskaTransonicDiskAccretion1982, 
MuchotrzebTransonicAccretionFlow1982,
MuchotrzebTransonicAccretionFlow1983, 
HoshiBasicPropertiesStationary1977, 
MatsumotoViscousTransonicFlow1984, 
AbramowiczAccretionDisksKerr1997}. 
Multiple sonic points implied multiple solutions --- connected via steady shocks
\cite{FukueTransonicDiskAccretion1987, 
ChakrabartiStandingShocksIsothermal1989, 
ChakrabartiStandingShocksIsothermal1990, 
ChakrabartiStandingRankineHugoniotShocks1989, 
AbramowiczStandingShocksAdiabatic1990, 
ChakrabartiSmoothedParticleHydrodynamics1993, 
CaditzAdiabaticShocksAccretion1998}. 
Subsequent work sought to resolve this multiplicity of solutions through stability analysis.
The first bounds on shock wave stability lead to a simple condition: inner shocks are unstable, and outer shocks are stable \cite{NakayamaHydrodynamicInstabilityAccretion1992}.
Following papers considered more general accretion disk models \cite{NakayamaHydrodynamicInstabilityAccretion1993,
NakayamaDynamicalInstabilityStanding1994, 
NakayamaUnstableStandingShock1996,
YangShockStudyFully1995, 
ChakrabartiGlobalSolutionsViscous1996, 
YuanSonicPointsShocks1996, 
FukumuraIsothermalShockFormation2004}, 
nonaxisymmetric perturbations \cite{GuNonaxisymmetricInstabilitiesShocked2003, 
GuNonaxisymmetricInstabilitiesShocked2006, 
NagakuraGeneralRelativisticHydrodynamic2008, 
LeStandingShockInstability2016}, 
and additional thermodynamic and electromagnetic effects 
\cite{KumarDissipativeAdvectiveAccretion2014, 
SahaModelDependenceMultitransonic2016, 
SarkarPropertiesMagneticallySupported2018}.

All previous analyses assumed Rankine-Hugoniot shock conditions. 
But skipping straight to zero viscosity neglects important details like angular momentum dissipation.
Multiple scales matched asymptotic analysis, {in the limit} of vanishing viscosity, provides a straightforward and comprehensive procedure to determine all mathematical properties of accretion disk shocks.
Such tools have been used to great effect in accretion disk boundary layers
\cite{RegevAsymptoticModelsAccretion1995}, 
and reduced model derivation throughout astrophysical fluid dynamics \cite{JulienMagnetorotationalInstabilityRecent2010}.
Geometric Singular Perturbation Theory provides a rigorous justification for the asymptotic analysis \cite{SzmolyanCanardsR32001,MitryFoldedSaddlesFaux2017}, and has been applied to astrophysical shocks in stellar winds \cite{CarterTransonicCanardsStellar2017,BauerExistenceTransonicSolutions2021}.

We revisit classical models of thin accretion disks using the tools of multiple scale matched asymptotics.
In \cref{sec:theory} we present the techniques of the current investigation.
We start with the equation derivation in \cref{sec:equations}, followed by an outline of the asymptotic analysis procedure in \cref{sec:asymptotics}, and symbolic and numerical computational tools in \cref{sec:computational-methods}.
In \cref{sec:results} we present the results of our investigation.
\Cref{sec:steady-shocks} categorises the possible steady solutions to isothermal accretion disks, and \cref{sec:stability} derives analytical expressions for the growth/decay rates of these solutions.
\Cref{sec:steady-asymptotics,sec:smooth-supersonic,sec:stability-asymptotics} detail the analysis step-by-step.
We then discuss the scientific implications of our findings in \cref{sec:discussion}, before concluding in \cref{sec:conclusion}.

\section{Theory}
\label{sec:theory}

\subsection{Equations of motion}
\label{sec:equations}
We begin with conservation of mass and momentum
	\begin{align}
    &\pd_t \rho + \nabla \cdot (\rho u ) = 0,\\
	&\pd_t u + u \cdot \nabla u + \frac{1}{\rho}\nabla \cdot ((p - \mu_1 \nabla \cdot u) I - \mu_2 (\nabla u + \nabla u^\top)) + \nabla \phi = 0.
    \end{align}
The viscous thin-disk equations are derived using the following assumptions
	\begin{enumerate}
	\item Cylindrical coordinates $(r,\theta,z)$.
	\item Axisymmetric flow $\pd_\theta \to 0$.
	\item Isothermal equation of state $p = K \rho$.
	\item Constant kinematic diffusivities $\nu_1 = \mu_1/\rho, \nu_2 = \mu_2/\rho$.
	\item Monatomic gas, with bulk viscosity $\nu_1 = 0$.
	\item No self-gravity.
	\item Pseudo-Newtonian black hole potential $\phi = \frac{GM}{r-r_{Sc}}$.
	\item Thin disk $z/r \ll 1$.
	\item Hydrostatic balance $\pd_z u = 0$.
	\end{enumerate}
The horizon radius in the Pseudo-Newtonian potential \cite{PaczynskyThickAccretionDisks1980} is given by the Schwarzschild radius $r_{Sc} = \frac{2 GM}{c^2}$, where $G$ is the gravitational constant, $M$ is the black hole mass, and $c$ is the speed of light.

\subsubsection{Non-dimensional equations}
We nondimensionalise according to length and time scales given by the sound speed $\sqrt{K}$, defining the dimensional velocity scales $U_r, U_\theta$, length scale $R$, and time scale $T$
	\begin{align}
    U_r &= U_\theta = \sqrt{K}, &
    R &= \frac{GM}{2K}, &
    T &= \frac{R}{U_r},
	\end{align}
which reduces the equations to 
	\begin{equation}\label{eq:delta}
    \pd_t \delta + u \pd_r \delta + \pd_r u + \frac{u}{r} = 0,
    \end{equation}
    \begin{align}\label{eq:u}
    \pd_t u + u\pd_r u + \pd_r \delta + \br{\frac{2}{(r-r_h)^2} - \frac{\ell^2}{r^3}} &= \nonumber\\
         2\eps \left(\pd_r^2 u + \frac{\pd_r u}{r} \right. &-\left. \frac{u}{r^2} + \pd_r \delta \pd_r u\right),
    \end{align}
    \begin{equation}\label{eq:l}
    \pd_t \ell + u \pd_r \ell =
    \eps \br{\pd_r^2 \ell - \frac{\pd_r \ell}{r} + \br{\pd_r \ell - \frac{2\ell}{r}}\pd_r \delta }.
	\end{equation}
where we have defined the dimensionless horizon scale $r_h$ and inverse Reynolds number $\eps$
	\begin{align}
    r_h &\equiv \frac{4K}{c^2}, &
%    \eps &\equiv \frac{\nu_2 }{U_r R},
    \eps &\equiv \frac{2 \nu_2 \sqrt{K}}{G M},
	\end{align}
and rewritten in terms of the logarithmic density $\delta$ and specific angular momentum $\ell$
    \begin{align}
    \delta(t, r) &= \log \rho(t, r), &
    \ell(t, r) &= r u_\theta(t, r).
    \end{align}

\subsection{Asymptotic analysis}
\label{sec:asymptotics}
\Cref{eq:delta,eq:u,eq:l} are singularly perturbed in $\eps$.
Neglecting viscous terms fails within narrow shock boundary layers of thickness $O(\eps)$.
In these regions higher-order derivatives must be included.

\subsubsection{Multiple-scales matched asymptotic expansions}
We analyse these shocks using multiple scales matched asymptotic expansions in $\eps$.
We divide the analysis into two subproblems.
The \emph{reduced/outer problem} determines the `outer' fluid variables $f^-(r)$ and $f^+(r)$ for radii less than ($r<r_s$) and greater than ($r>r_s$) the shock radius $r_s$ respectively.
(Here $f$ represents an arbitrary fluid variable, encompassing the choices $\delta, u, \ell$.)
The \emph{layer/inner} problem solves the `inner' fluid variables $f(x)$ within the shock itself by defining rescaled inner coordinates $x = \frac{r - r_s}{\eps}$.
Applying this coordinate transformation to \cref{eq:delta,eq:u,eq:l} rescales the radial derivatives
	\begin{align}
    r &= r_s + \eps x, &
    \pd_r &\to \frac{1}{\eps}\pd_x.
	\end{align}
For each problem, we then expand the fluid variables $f$ in a formal asymptotic series in $\eps$, 
	\begin{align}
    f^\pm(r) &= \sum_{k=0}\eps^k f^\pm_k(r), &
    f(x) &= \sum_{k=0}\eps^k f_k(x),
	\end{align}
and solve the problems order-by-order in $\eps$.

\subsubsection{Asymptotic matching} 
The final stage connects inner and outer subproblems through boundary conditions.
We specify asymptotic agreement in the region $x \sim \eps^{-1/2}$ in the limit $\eps\to 0$,
    \begin{align}
	\lim_{\eps\to 0} f(\pm\eps^{-1/2}x) &\sim \lim_{\eps\to 0} f^{\pm}(r_s\pm\eps^{+1/2}\,x ).
    \end{align}
This limit simultaneously allows the layer coordinate $x$ to approach infinity and the reduced coordinate $r$ to approach $r_s$.
We derive order-by-order boundary conditions by expanding each variable in an asymptotic series in $\eps$,
expanding each term of the outer variable $f^\pm(r)$ as a Taylor series around $r_s$,
and equating each order in $\eps$
    \begin{align}
    \lim_{x\to\pm\infty} \sum_{k=0} \eps^k f_k(x)
    \sim \lim_{x\to\pm\infty}  \sum_{k=0}\eps^k\br{\sum_{\ell=0}^k\frac{x^{\ell}}{\ell!} \pd_r^\ell f^{\pm}_{k-\ell}(r_s)}.
    \end{align}
The first two conditions simplify to
	\begin{align}
	\lim_{x\to\pm\infty}f_0(x)\sim f^\pm_0(r_s), \quad
	\lim_{x\to\pm\infty}f_1(x)\sim f^\pm_1(r_s) + \pd_r f^\pm_0(r_s)\, x.
	\end{align}
This procedure allows us to determine analytic formulae for the asymptotic behaviour of the steady problem and growth/decay rates of the shock stability problem.

\subsection{Computational methods}
\label{sec:computational-methods}
All asymptotic calculations are verified in a Mathematica script available at \href{https://github.com/ericwhester/isothermal-accretion-disk-shocks}{github.com/ericwhester/isothermal-accretion-disk-shocks}.
We also validate our analysis with simulations using the flexible spectral code Dedalus \cite{BurnsDedalusFlexibleFramework2020}.
The Dedalus code is available from \href{dedalus-project.org}{https://dedalus-project.org/},
and the scripts used to run and plot the analysis in this paper are available at the same \href{https://github.com/ericwhester/isothermal-accretion-disk-shocks}{github repository}.

\section{Results}
\label{sec:results}

\subsection{Steady transonic solutions}
\label{sec:steady-shocks}
We find stationary solutions by neglecting time derivatives in \cref{eq:delta,eq:u,eq:l}.
The steady mass conservation law \cref{eq:delta} immediately determines the logarithmic density in terms of the radial velocity
\begin{equation}\label{eq:steady-delta}
\delta(r) = -\log (- r u).
\end{equation}
We then solve the steady momentum equations order-by-order in $\eps$.

\subsubsection{Conserved quantities for inviscid flow}
Substituting the steady log-density (\cref{eq:steady-delta}) into the steady radial momentum equation \cref{eq:u}, and solving the asymptotic expansions of \cref{eq:u,eq:l}, we find that the leading asymptotic equations for the reduced problem in $u$ and $\ell$ simplify to conservation laws for the specific energy $e$ and specific angular momentum $\ell$,
    \begin{align}\label{eq:steady-zeroth-outer}
    \pd_r e &= 0,&
    \pd_r \ell_0^\pm &= 0,
    \end{align}
where the specific energy $e$ is given by
	\begin{align}
    e &\equiv {\frac{1}{2}\br{{u^\pm_{r,0}}^2 + \frac{{\ell_0^\pm}^2}{r^2}} - \frac{2}{r-r_h} - \log ( r |u^\pm_{r,0}|) }.
	\end{align}

\subsubsection{Shock boundary conditions}
The inviscid solutions (\cref{eq:steady-zeroth-outer}) are not valid across a shock.
To derive the shock jump conditions we rescale to the shock length scale, and solve the leading order asymptotic problem in $u_0(x), \ell_0(x)$, detailed in \cref{sec:steady-asymptotics}.
The solutions are given by
	\begin{align}
    u_0(x) &= -\cosh a - \sinh a \tanh\br{\frac{\sinh (a)}{2} \, (x - x_0)}, \\
    \ell_0(x) &= \ell^\pm_0(r_s) \equiv \ell.
	\end{align}
where we have defined $a \equiv \log |u^+_0(r_s)|>0$.
The leading order specific angular momentum is therefore conserved across a shock, and so is constant throughout the domain.
We define this constant as $\ell \equiv \ell_0(x) = \ell_0^\pm(r)$.
	
\textbf{Sonic points} ---
The outer radial velocity $u^\pm_0$ is given by level sets of the specific energy $e$.
The global behaviour of $u^\pm_0$ therefore depends on the critical points of the specific energy $e$.
Partial derivatives of $e$ with respect to the velocity $u^\pm_0$ and radius $r$ give the critical point conditions
	\begin{align}
    \pd_{u^\pm_0} e &= u^\pm_0-\frac{1}{u^\pm_0} = 0, &
    \pd_{r} e &= \frac{1}{r} - \frac{2}{(r-r_h)^2} + \frac{\ell^2}{r^3} = 0.
	\end{align}
The critical points $(r_*, u^\pm_{0,*})$ are therefore given by $u^\pm_{0,*} = -1$, and 
	\begin{align}
	\label{eq:critical-conditions}
    r_*^4 - 2(1+r_h) r_*^3 + (\ell^2 +r_h^2) r_*^2 - 2 \ell^2 r_h r_* + \ell^2 r_h^2 &= 0.
	\end{align}
There are at most four critical points of $e$.
Proceeding inwards, the first three are a saddle type critical point $r_{*,2}$, a centre type critical point $r_{*,0}$, and another saddle type critical point $r_{*,1}$.
The final unphysical critical point occurs within the black hole for $r < r_h$.
The only locations where the flow can become transonic are the inner $r_{*,1}$ and outer $r_{*,2}$ sonic points.	
Geometric Singular Perturbation theory proves that the transonic flow through the sonic points (canard points in the GSPT literature) persists even with added viscosity \cite{SzmolyanCanardsR32001,CarterTransonicCanardsStellar2017,MitryFoldedSaddlesFaux2017}.

\begin{figure}
	\centering
    \includegraphics[width=\linewidth]{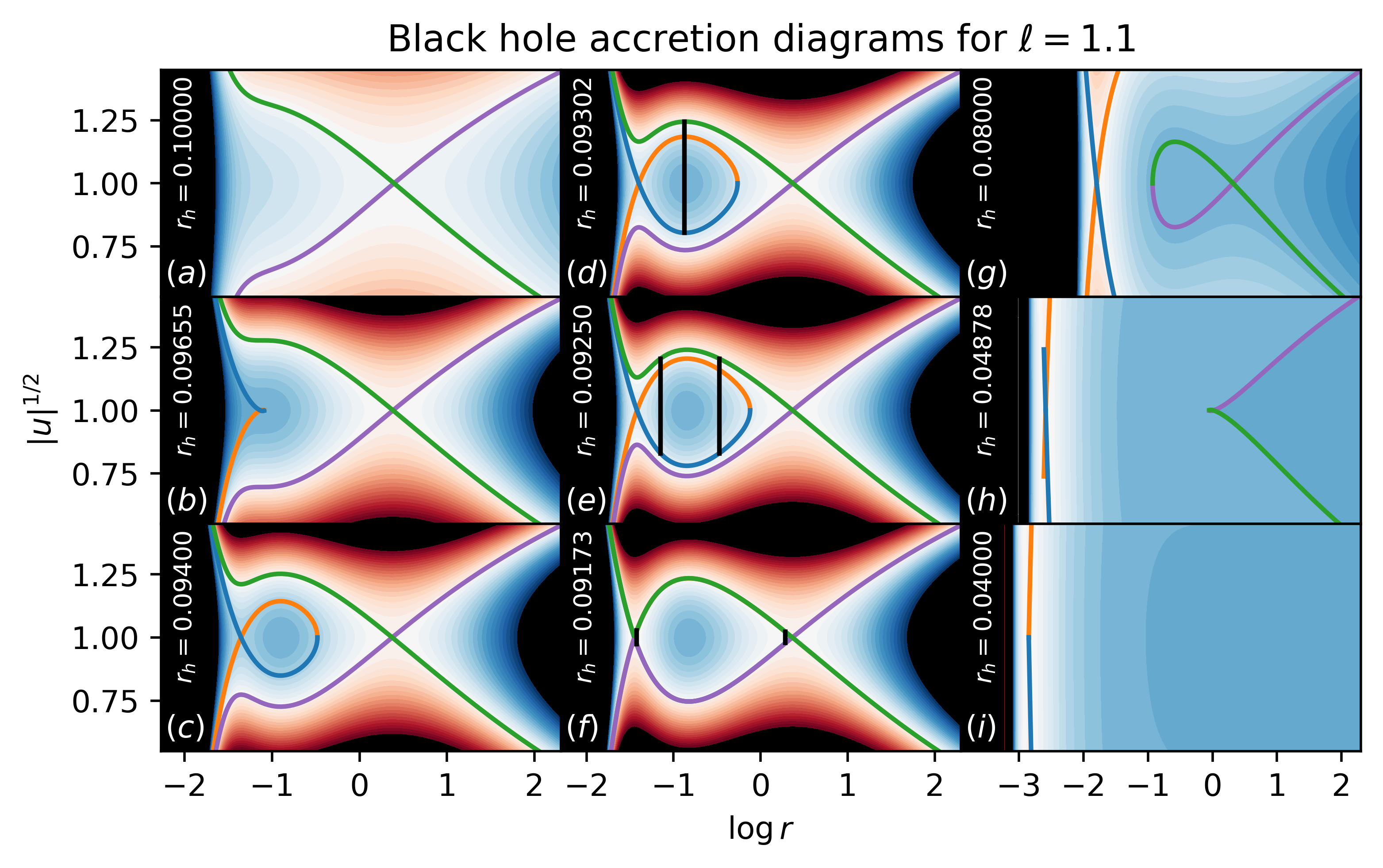}
    \caption{Steady black hole accretion diagrams for varying horizon radius $r_h$ at angular momentum $\ell = 1.1$. 
    The background colour plots level sets of the specific energy $e$, centred about the highest energy sonic point.
    Transonic trajectories through the inner ($u^-_0$, blue, orange) and outer ($u^+_0$ green, purple) sonic points are plotted for the nine different accretion regimes mentioned above. 
    Zero, one, and two shocks (black lines) are possible.
    Transonic accretion solutions must connect $\lim_{r\to\infty} u = 0$ to $\lim_{r\to r_h} u \to \infty$.}
    \label{fig:black-hole-accretion-diagrams}
\end{figure}

\textbf{Shock and sonic point regimes} --- 
The zeroth order layer problem shows that transonic shocks occur where the velocity jumps from $u^+_0$ to $1/u^+_0 = u^-_0$ --- i.e. the Rankine-Hugoniot jump condition.
If we further require outer boundary conditions that are subsonic as $r \to \infty$ but supersonic as $r \to r_h$, then the shocks must also connect the trajectories through the inner $r_{*,1}$ and outer $r_{*,2}$ sonic points.
We thus determine the shock location by finding intersections of the projection of the outer transonic velocity $1/u^+_0(r)$ with the inner transonic velocity $u^-_0(r)$.
This completely describes the structure of the solution as a function of $\ell$ and $r_h$.
We plot these regimes in \cref{fig:black-hole-accretion-regime-diagram}, and summarise them in order of decreasing $r_h$ in \cref{tab:accretion-regimes}, giving example plots in \cref{fig:black-hole-accretion-diagrams}:

\begin{figure}
	\centering
    \includegraphics[width=.8\linewidth]{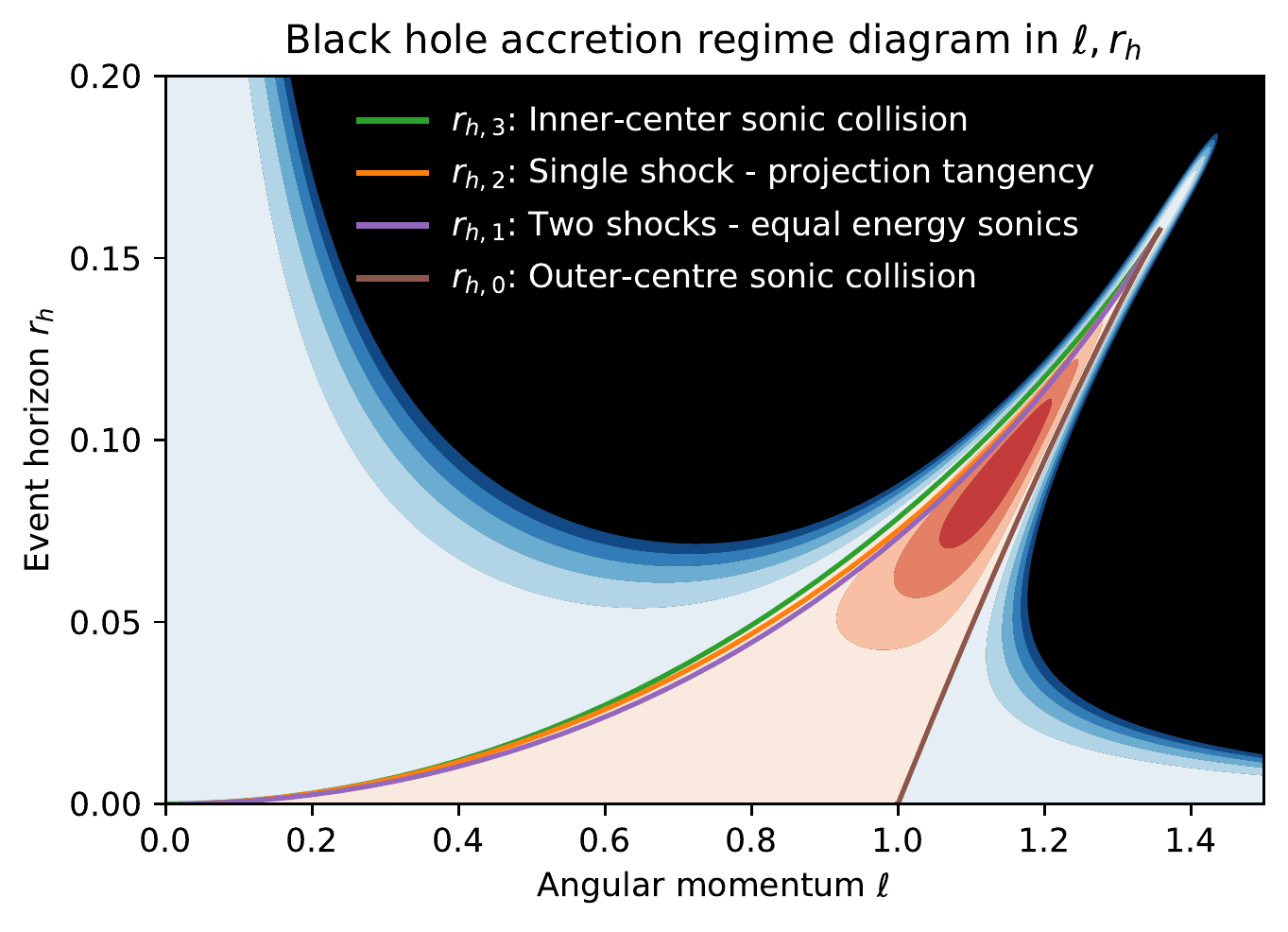}
    \caption{
    Regime diagram for black hole accretion.
    The colour is given by the determinant of the polynomial in \cref{eq:critical-conditions}.
    Two sonic points (saddle type) and a centre type critical point exist in the red region.
    A single sonic point exists in the blue region.
    Two shocks (and therefore three transonic solutions) are possible between the purple $r_{h, 1}$ and orange $r_{h,2}$ curves.
    One transonic solution exists otherwise.}
    \label{fig:black-hole-accretion-regime-diagram}
\end{figure}

The leading order inner problem is translation symmetric.
The boundary conditions cannot determine the location of $u_0$ (the value of $x_0$).
To do so, we proceed to the next order of the reduced and layer problems.

\begin{table}
\centering
  \begin{tabular}{llcl}
    Figure & Regime & Shocks & Description\\
    \hline
	2 $(a)$ &
	$r_{h,3} < r_h$ &
	0 &
	Transonic flow through outer sonic. \\
	2 $(b)$ &
	$r_h = r_{h,3}$ &
	0 &
	Inner-centre sonic collision.\\
	2 ($c$) &
	$r_{h,2} < r_h < r_{h,3}$ &
	0 &
	No transonic shock connection. \\
	2 $(d)$ &
	$r_h = r_{h,2}$ & 
	1 &
	Single transonic shock. \\
	2 $(e)$ & 
	$r_{h,1} < r_h < r_{h,2}$ &
	2 &
	Two possible shocks.\\
	2 $(f)$ &
	$r_h = r_{h,1}$ & 
	2 &
	Equal energy sonic points. \\
	2 $(g)$ &
	$r_{h,0} < r_h < r_{h,1}$ &
	0 &
	Inner sonic point has higher energy. \\
	2 $(h)$ &
	$r_h = r_{h,0}$ &
	0 &
	Outer-centre sonic collision. \\
	2 $(i)$ &
	$r_h < r_{h,0}$ & 
	0 &
	Transonic flow through inner sonic.
    \end{tabular}
  \caption{Table of accretion flow shock regimes.}
  \label{tab:accretion-regimes}
\end{table}

\subsubsection{First order outer solution}
The first order outer problem for $u_1^\pm, \ell_1^\pm$ is linear (\cref{sec:steady-asymptotics}), and the specific angular momentum perturbation is found to be
    \begin{align}
    \ell_1^\pm &= 2\ell(\rho_0^\pm - \rho_\infty).
    \end{align}

\subsubsection{First order inner solution}
The first order layer problem for $u_1$ is linear, and implies
    \begin{align}
    u_1(x) &= g(r_s) u_{1,i}(x) + b_1 u_{1,1}(x) + b_2 u_{1,2}(x),
    \end{align}
where
    \begin{align}
    g(r_s) &= -\frac{1}{2}\br{\frac{1}{r_s} - \frac{2}{(r_s - r_h)^2}+ \frac{\ell^2}{r_s^3}},\nonumber\\
    u_{1,1}(x) &= u_0'(x),\nonumber\\
    u_{1,2}(x) &= u_0(x) + x u_0'(x) + \sech(a),\nonumber\\
    u_{1,i}(x) &= c_1(x) u_{1,1}(x) + c_2(x) u_{1,2}(x),\nonumber 
    \end{align}
and $c_1(x), c_2(x)$ are provided in \cref{sec:steady-asymptotics}.
The angular momentum is given by
    \begin{align}
    \ell_1(r) &= \frac{\ell}{r_h}\br{\sech a + \cosh(a - x\sinh a) } \sech\br{\frac{\sinh a}{2}x }^2 + d_1,
    \end{align}
where $d_1$ ensures matching conditions between $\ell_1^\pm$.

\subsubsection{Numerical simulation}
\label{sec:steady-shock-numerics}
\begin{figure}
	\centering
    \includegraphics[width=\linewidth]{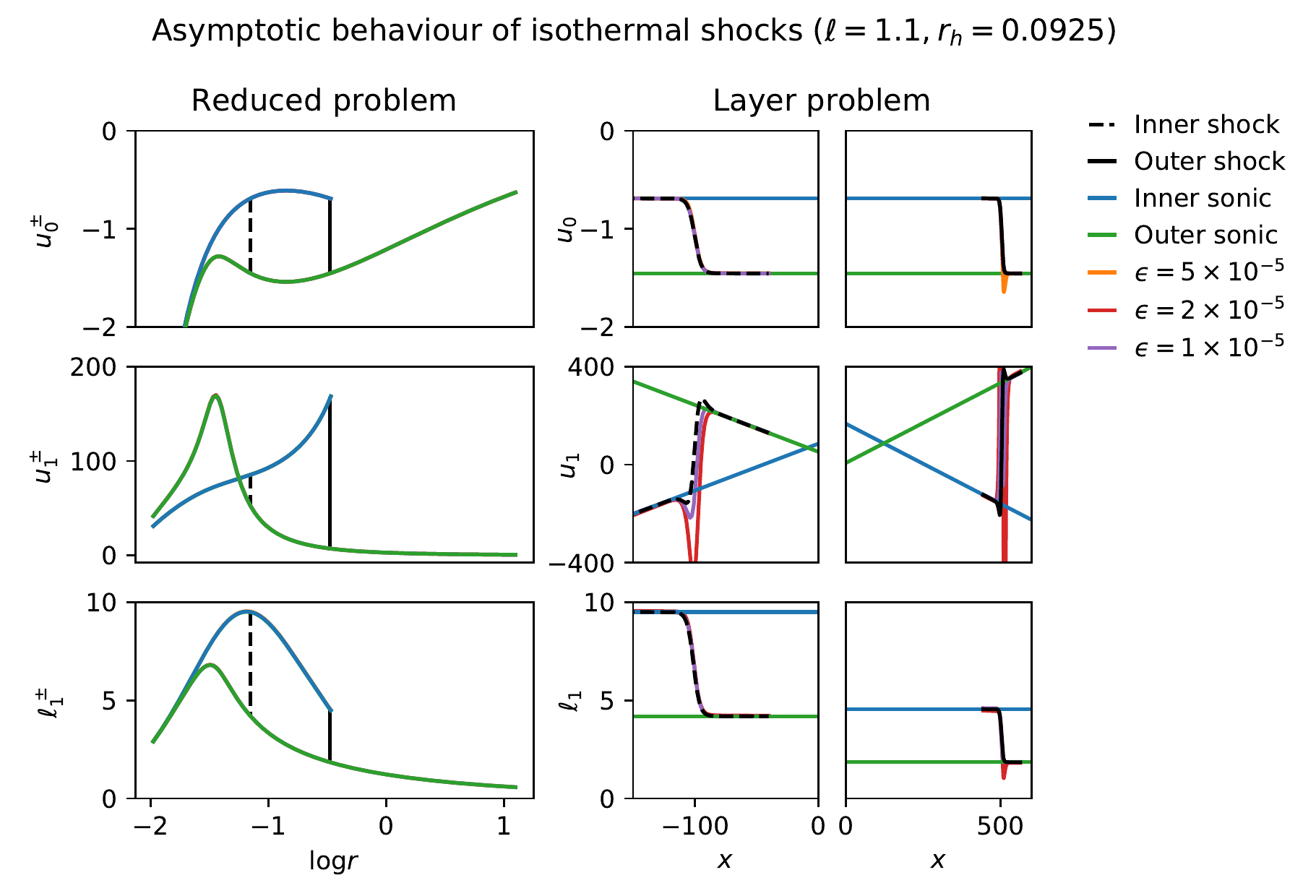}
    \caption{Profiles of each asymptotic expansion at $\ell = 1.1, r_h = 0.0925$.
    The left column plots the zeroth order outer velocities $u^\pm_0$, the first order outer velocities $u^\pm_1$, and the first order angular momentum $\ell^\pm_1$ (top, middle, bottom respectively), for the flows through the inner (blue) and outer (green) transonic points.
    Shocks are plotted in black (inner dashed, outer solid).
    The second and third columns plot the zeroth order velocity $u_0$, first order velocity $u_1$, and first order angular momentum $\ell_1$ through the inner and outer shock respectively.
    The second and third columns also plot the corresponding asymptotic behaviour of the left (blue) and right (green) outer solutions as they approach the corresponding shock.
    Each figure plots Richardson extrapolants calculated from the nonlinear simulations for different values of $\eps$ (legend).
    The empirical calculations are indistinguishable from the asymptotic calculations for the reduced problem.
    Empirical convergence is slower for the shocks because of the unconstrained $u_{1,1}$ term, but the shock offset $x_0$ and limiting behaviour as $x\to\pm\infty$ are correct.
    }
    \label{fig:black-hole-asymptotics-summary}
\end{figure}

We empirically validate asymptotic convergence of the solution as $\eps \to 0$ for ${\ell = 1.1}, {r_h = 0.0925}$ using the spectral code Dedalus.
We solve the fully nonlinear steady equations \cref{eq:delta,eq:u,eq:l}
and compare the nonlinear solution with the asymptotic solutions determined above.
We also specify Dirichlet boundary conditions on the velocity at $r_0 = 1.5 \, r_h$ and $r_1 = 3$ with boundary conditions determined by the first order asymptotic expansions
	\begin{align}
	u(r_0) &= u^-_0(r_0) + \eps u^-_1(r_0), &
	u(r_1) &= u^+_0(r_1) + \eps u^+_1(r_1), \nonumber\\
	\ell(r_0) &= \ell + \eps \ell^-_1(r_0), &
	\ell(r_1) &= \ell + \eps \ell^+_1(r_1), \nonumber\\
	& &
	\delta(r_1) &= -\log(-r_1 u(r_1)).
	\end{align}
These equations are solved numerically as a nonlinear boundary value problem, which uses Newton iteration to converge to a solution of the equations.
To ease numerical costs, we use a spectral element method, which divides the domain into two outer regions and an inner region with boundaries $(r_0, r_s + \eps (x_0 - 100), r_s + \eps (x_0+100),r_1)$.
We solve this equation for $\eps = \num{1e-4}, \num{5e-5}, \num{2e-5}, \num{1e-5}$.
We calculate empirical asymptotic convergence using Richardson extrapolation in $\eps$.
That is, given a power series in $\eps$, $y(x, \eps) = \sum_{k=0} \eps^k y_k(x)$ for two values of $\eps$, we apply the following linear combination to cancel the first order term $y_1$, 
	\begin{align}
	y_0(x) &= \frac{\frac{y(x,\eps_1)}{\eps_1} - \frac{y(x,\eps_2)}{\eps_2}}{\eps_1^{-1} - \eps_2^{-1}} + \O(\eps^2),
	\end{align}
giving a second order accurate estimate of $y_0(x)$.	
Repeating the process on $(y - y_0)/\eps$ allows an estimate of $y_1$, and so on.
We summarise all analytic and empirical asymptotic behaviour for the reduced and layer problems  up to and including first order in \cref{fig:black-hole-asymptotics-summary}.
The empirical solver exactly reproduces the asymptotic calculations for the outer scale, and shows clear convergence as $\eps \to 0$ for the shocks.
The only disagreement between the calculations is due to the unconstrained $u_{1,1}$ term in the shock problem.
We thus conclude that our asymptotic calculations are correct.

\subsection{Stability analysis}
\label{sec:stability}
We have shown that multiple steady accretion solutions exist for some range of $r_h, \ell$.
To determine which of these solutions are physical, we complete a linear stability analysis.
\subsubsection{Asymptotic analysis of linear stability}
Each fluid variable $f(r,t)$ (including $\delta, u,\ell$) separates into steady and unsteady components $f(r, t) = {f}(r) + \tilde{f}(r,t)$,
where ${u}, {\delta}, {\ell}$ are given by the steady solutions calculated in the previous section.
Substituting into \cref{eq:delta,eq:u,eq:l} gives
	\begin{align}\label{eq:unsteady-nonlinear-delta}
	&\pd_t \tilde{\delta} 
	+\tilde{u} \pd_r {}{\delta}
	+{}{u}   \pd_r\tilde{\delta}
	+\pd_r\tilde{u}	
	+\frac{\tilde{u}  }{r} = -\tilde u \pd_r \tilde \delta,\\
	&\pd_t\tilde{u} \label{eq:unsteady-nonlinear-u}
	+\tilde{u} \pd_r {}{u}
	+{}{u} \pd_r \tilde{u}
	+\pd_r\tilde{\delta }
	-\frac{2 {}{\ell } \tilde{\ell }}{r^3} -2\eps \bigg(
		\pd_r^2\tilde{u}
		+\frac{\pd_r \tilde{u}}{r}
		-\frac{\tilde{u}}{r^2} \nonumber\\
		&\qquad\qquad+\pd_r{}{u} \pd_r\tilde{\delta }
		+\pd_r{}{\delta } \pd_r\tilde{u}
		\bigg) = 
		-\tilde u \pd_r \tilde u + \frac{\tilde \ell^2}{r^3} + \eps\pd_r \tilde u \pd_r \tilde \delta,\\
	&\pd_t\tilde{\ell } \label{eq:unsteady-nonlinear-l}
	+\tilde{u} \pd_r{}{\ell }
	+{}{u} \pd_r\tilde{\ell }
	-\eps  \bigg(
		\pd_r^2 \tilde{\ell }
		-\frac{\pd_r\tilde{\ell }}{r}
		+\pd_r{}{\delta } \pd_r\tilde{\ell }
		+\pd_r{}{\ell } \pd_r\tilde{\delta }\nonumber\\
		&\qquad\qquad-2\frac{ \tilde{\ell } \pd_r {}{\delta } + {}{\ell } \pd_r\tilde{\delta }}{r}\bigg)= 
		-\tilde u \pd_r \tilde \ell + \eps \bigg( \pd_r \tilde \ell - \frac{2 \tilde \ell}{r} \bigg) \pd_r \tilde \delta .
	\end{align}
We then linearise \cref{eq:unsteady-nonlinear-delta,eq:unsteady-nonlinear-u,eq:unsteady-nonlinear-l} about the steady states by discarding the nonlinear right hand sides.
Each problem is then divided into inner and outer regions, where we express each steady ${}{f}$ and unsteady $\tilde{f}$ variable as a formal asymptotic series, as in \cref{sec:steady-shocks}.

\subsubsection{Smooth stability analysis}
\label{sec:smooth-stability}
\textbf{Subsonic regime, $r> r_*$} --- 
We derive leading order asymptotic behaviour in the subsonic regime by setting $\eps \to 0$.
We simplify by using ${}{\delta}_0^+ = - \log (- r {}{u}^+_0)$, and ${}{\ell}_0^+ = \ell$,
and abbreviate by dropping $\pm$ superscripts and $0$ subscripts,
	\begin{align}
	(\pd_t +{} u \pd_r)  \tilde \delta 
	+ \br{\pd_r - \frac{{} u'}{{} u}} \tilde u &= 0,\\
	\pd_r \tilde \delta
	+ \br{\pd_t + {} u' + {} u \pd_r }\tilde u
	-\frac{2 \ell }{r^3} \tilde \ell &= 0,\\
	(\pd_t + {} u \pd_r) \tilde \ell &= 0.
	\end{align}
The specific angular momentum $\tilde \ell$ is advected by the background flow.
If initially $\tilde \ell(0,r) \to 0$ as $r \to \infty$, then all $\tilde \ell$ will eventually be advected through the sonic point.
After this point, disturbances evolve without any perturbation in specific angular momentum.
For this reason, we can consider stability in the absence of angular momentum perturbation $\tilde \ell$ for long times.
Hence, we deal with sound waves alone.
Rearranging, we find
	\begin{align}
    \pd_t^2 \tilde u + \pd_r\br{2{} u\pd_t \tilde u + {} u\pd_r\br{\br{{} u - \frac{1}{{} u}}\tilde u} } &= 0.
	\end{align}
To prove stability, we derive a decreasing energy functional for the system.
We define
	\begin{align}\label{eq:perturbation-energy}
    &E = \int_{r_*}^\infty \frac{1}{2} \br{h(r) (\pd_t \tilde u) ^2 - {} u \pd_r\br{h(r) \tilde u}^2 }dr,\\
    &\text{ where } h(r) = {} u(r) - \frac{1}{{} u(r)} > 0, \quad -1 < {} u < 0,\nonumber
	\end{align}
and observe that
	\begin{align}\label{eq:perturbation-energy-dissipation}
    \frac{d}{dt}E &= \sbr{-{} u h \br{\pd_r(h \tilde u) + \pd_t \tilde u}\pd_t \tilde u}_{r_*}^\infty  - \int_{r_*}^\infty  \frac{{} u'}{|{} u|} (\pd_t \tilde u)^2 dr.
	\end{align}
The boundary terms cancel, because $\tilde u$ tends to zero as $r \to \infty$, and $h(r_*) = -1 - 1/(-1) = 0$ at the sonic point $r_*$.
The integrand on the right hand side is manifestly positive.
Thus, total energy of linear perturbations can only decay over time beyond the outermost sonic point (though not necessarily to zero).

\textbf{Supersonic regime} --- 
Between the centre type critical point and the inner sonic point the background flow slows down, and the system energy increases (simply apply \cref{eq:perturbation-energy-dissipation} over $[r_h,r_*]$).
We cannot rely on an energy argument for stability.
Instead, we examine the normal modes $\tilde{u}(t,x) = e^{\lambda t}\tilde{u}(x)$
    \begin{align}
    \lambda^2 \tilde u + \pd_r\br{2{} u \lambda \tilde u + {} u\pd_r\br{\br{{} u - \frac{1}{{} u}}\tilde u} } &= 0.
    \end{align}
We transform this into a standard Sturm-Liouville form, following similar procedure to \cite{NakayamaHydrodynamicInstabilityAccretion1993}.
We first examine the integral $\psi \equiv \int \tilde u \,dr$, to solve the simpler equation
    \begin{align}
    \pd_r\br{\lambda^2 \psi + {}{u} \pd_r \br{2 \lambda \psi + h(r) \pd_r \psi}} = 0.
    \end{align}
Then, using a Liouville transformation $\psi(r(x)) = w(x) e^{\lambda x}$, we recover the equation
    \begin{align}
    \pd_x^2 w &= \frac{\lambda^2}{{}{u}(r(x))^2} w,
    \quad \text{where} \quad
    x(r) = \int_{r_h}^{r} \frac{- {}{u}}{{}{u}^2 - 1} dr'.
    \end{align}
The coordinates transform $r \in [r_h, r_*]$ to $x \in [0, \infty)$.

\begin{figure}
	\centering
    \includegraphics[width=.7\linewidth]{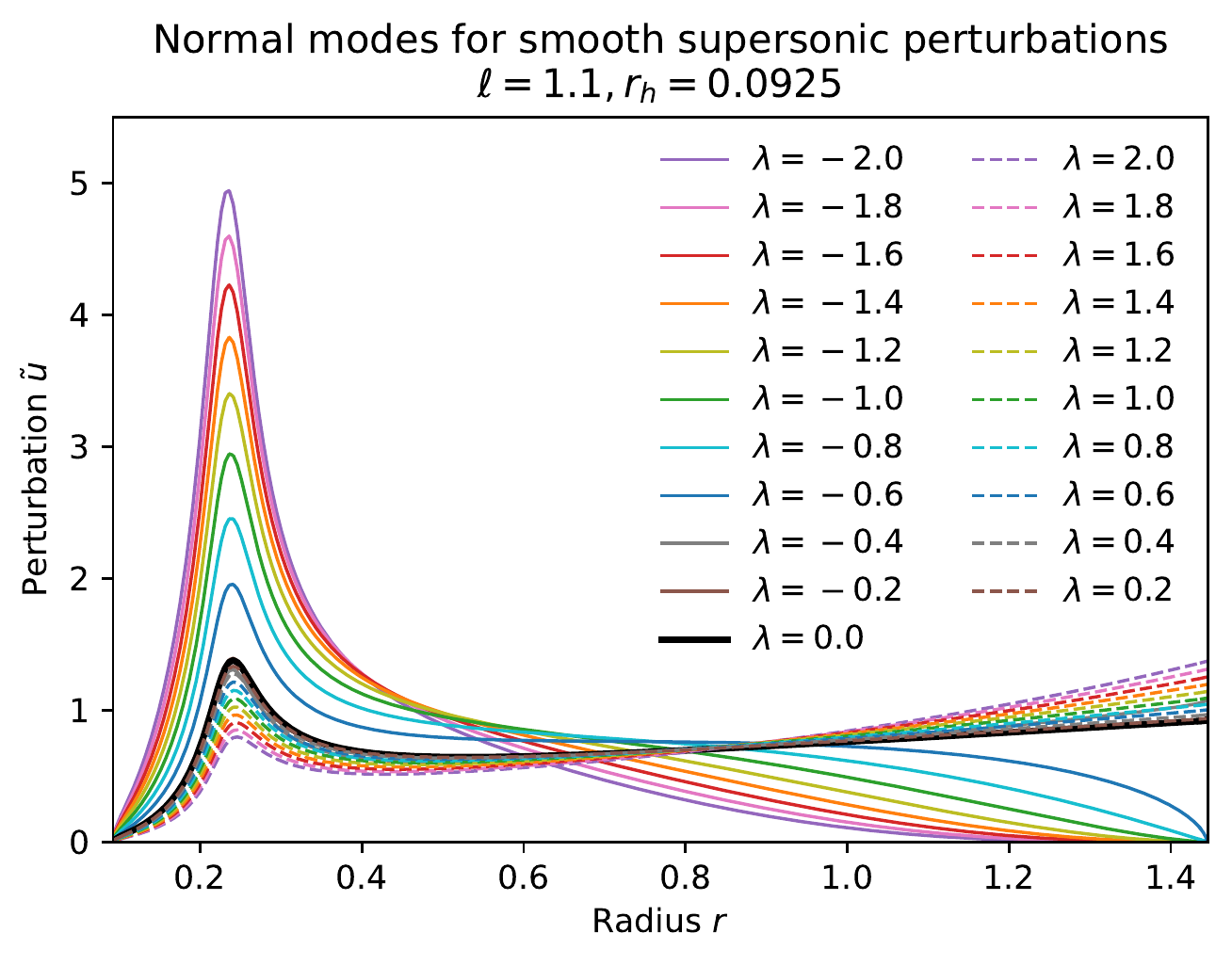}
    \caption{Normal mode profiles for supersonic perturbations ($r_h< r < r_*$) as a function of growth rate $\lambda$ (normalised so $\int_{r_h}^{r_*}\tilde{u} \, dr = 1$).
    Solutions with positive growth rate (dashed) have non-zero boundary values at the sonic point.
    All solutions with zero boundary values have negative growth rates (solid).
    Supersonic flow is therefore stable as the subsonic perturbations disappear at the sonic point as $t \to \infty$.
    }
    \label{fig:black-hole-smooth-pertubation-normal-modes}
\end{figure}

To prove stability, we show that normal modes with zero Dirichlet boundary conditions must have negative growth rates.
While the outer subsonic flow can temporarily perturb the supersonic region, eventually the flow decays to zero at the sonic point.
After this time, the supersonic flow must satisfy zero Dirichlet boundary conditions at $r_*$.
We show all such solutions decay over time using WKB asymptotic analysis near the sonic point (see \cref{sec:smooth-supersonic}).
In \cref{fig:black-hole-smooth-pertubation-normal-modes} we plot example normal modes for the supersonic region, and see that all modes that are zero at the sonic point have negative growth rates.

\begin{figure}[t]
	\centering
    \includegraphics[width=\linewidth]{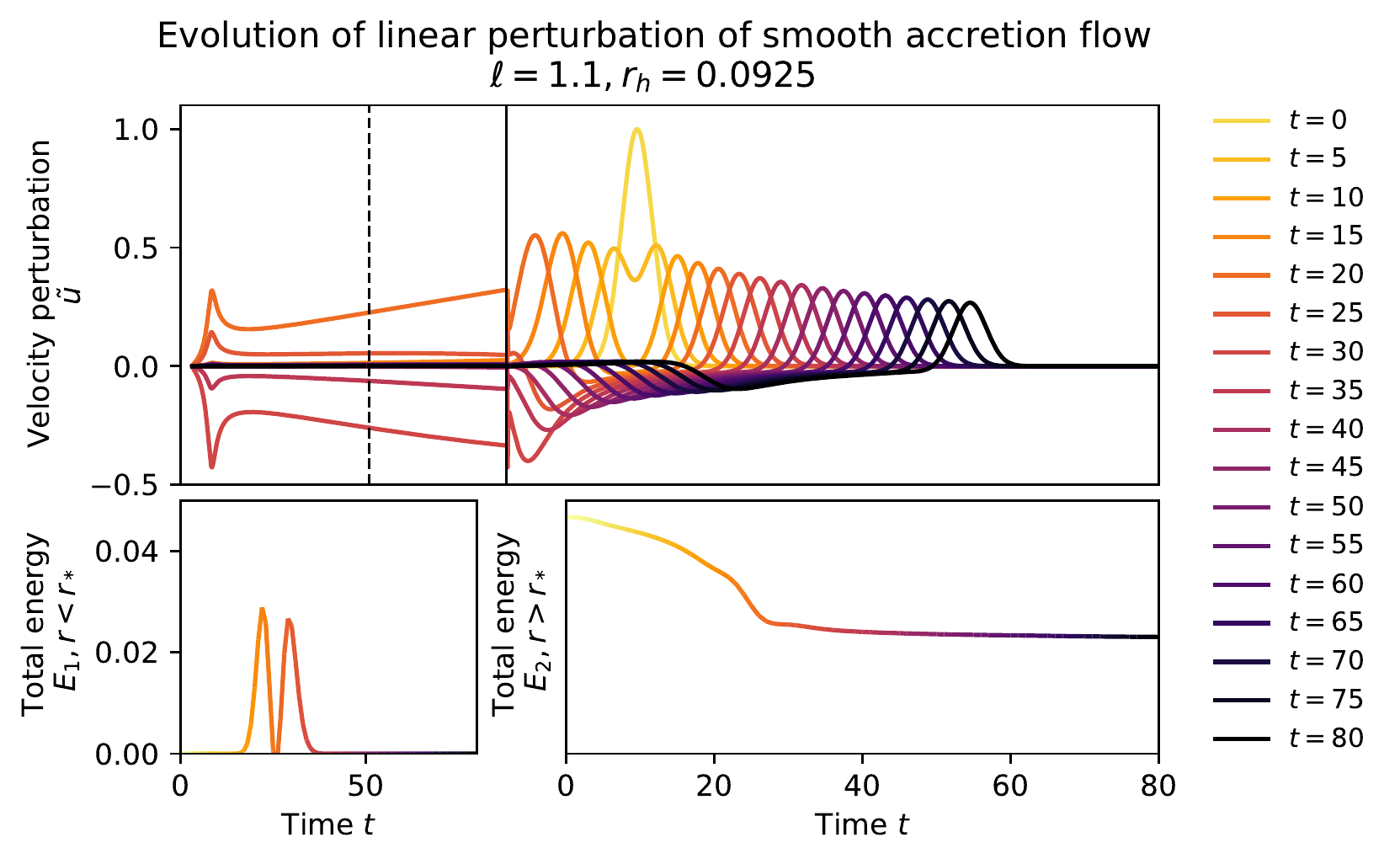}
    \caption{Evolution of linear perturbation of smooth steady flow at $\ell = 1.1, r_h = 0.0925$.
    The top row plots the velocity perturbation $\tilde{u}$ over space, colour-coded by time. 
    The left column zooms in on the flow near the black hole, and shows that the perturbations eventually decay within the (outer) sonic point (dashed line).
    The system at large $r$ tends to an outward propagating wave at unit speed.
    The total energy of the system beyond the sonic point (bottom right) strictly decreases over time, while the energy within the sonic point (bottom left) shows temporary increases that eventually decay to zero.
    }
    \label{fig:black-hole-linear-perturbation-smooth-evolution}
\end{figure}

We also simulate the evolution of an initial perturbation $\tilde{u}(0,x) = \exp\br{-((r-30)/5)^2}$ in \cref{fig:black-hole-linear-perturbation-smooth-evolution}.
The plot shows that the subsonic flow strictly decreases in energy over time (bottom right), with the system tending toward an outward-travelling wave as $t\to \infty$ (top right).
In contrast, the supersonic flow experiences a temporary increase in energy when the subsonic region injects energy (bottom left).
However, the inner energy eventually disappears, and the flow remains stable.

\subsubsection{Shock stability}
\begin{figure}
	\centering
    \includegraphics[width=\linewidth]{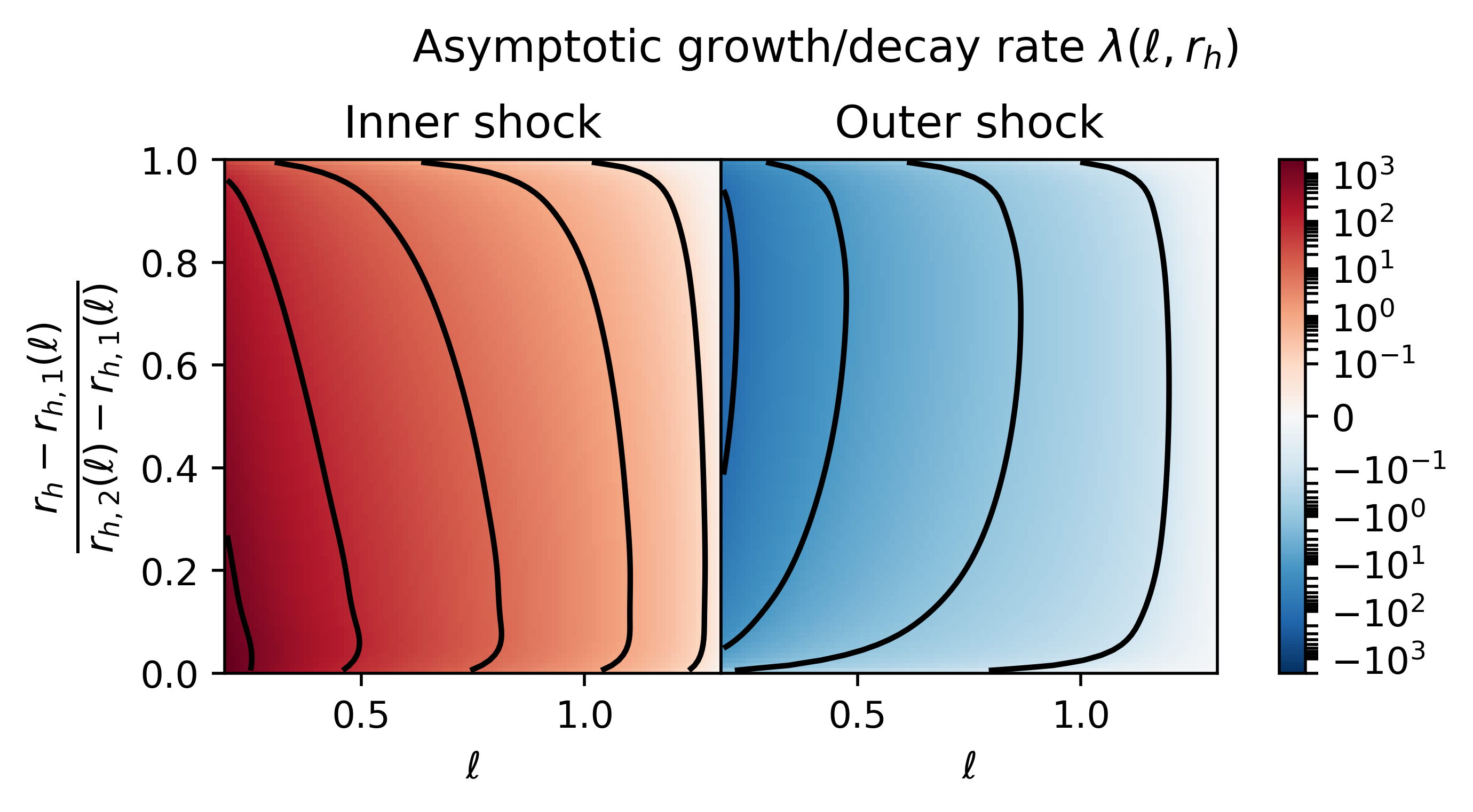}
    \caption{Growth/decay rate of inner/outer (left/right) shocks in terms of $\ell, r_h$.
    The vertical axis scales the distance between the $r_{h,1}(\ell)$ and $r_{h,2}(\ell)$ curves (\cref{fig:black-hole-accretion-regime-diagram}) to unity.
	The color bar is scaled logarithmically, with level sets shown at powers of 10.
	The inner shock growth rate is always positive (red), and the outer shock growth rate is always negative (blue).
    }
    \label{fig:black-hole-shock-stability-regimes}
\end{figure}

To analyse the shock stability problem, we must rescale to the shock size $\eps$.
The leading order problem shows
 	\begin{align}
	\tilde{u}_0(t,x) &= e^{\lambda_0 t} {}{u}_0'(x), &
	\tilde{\delta}_0(t,x) &= e^{\lambda_0 t} {}{\delta}_0'(x), &
	\tilde{\ell}_0(t,x) &= 0.
	\end{align}
To determine the growth rate $\lambda_0$ we apply a solvability condition on the next order problem (\cref{sec:stability-asymptotics}),
where we find
	\begin{align}
    \lambda_0 = -g(r_s) \sech a = \frac{\frac{1}{r_s}-\frac{2}{\left(r_s	-r_h\right){}^2} + \frac{\ell	^2}{r_s^3}}{u^+_0(r_s) + u^-_0(r_s)}.
	\end{align}
The inner and outer shock sizes are identical, so shock stability is determined purely by the sign of $-g(r_s)$, which is always positive for the inner shock, and negative for the outer shock.
We summarise the growth/decay rate $\lambda_0$ of shocks as a function of $\ell, r_h$ in \cref{fig:black-hole-shock-stability-regimes}.
The inner shock is always unstable (positive $\lambda_0$, red), and the outer shock is always stable (negative $\lambda_1$, blue).
The growth rate $\lambda_0$ tends to zero as the shocks tend to the sonic points (bottom), tend to each other (at the centre type critical point, top), and as the shocks become weaker (to the right).
Lower angular momentum black holes (left) lead to more extreme growth rates.

\subsubsection{Numerical validation of asymptotic shock stability}
\label{sec:unsteady-shock-numerics}
\begin{figure}
	\centering
    \includegraphics[width=\linewidth]{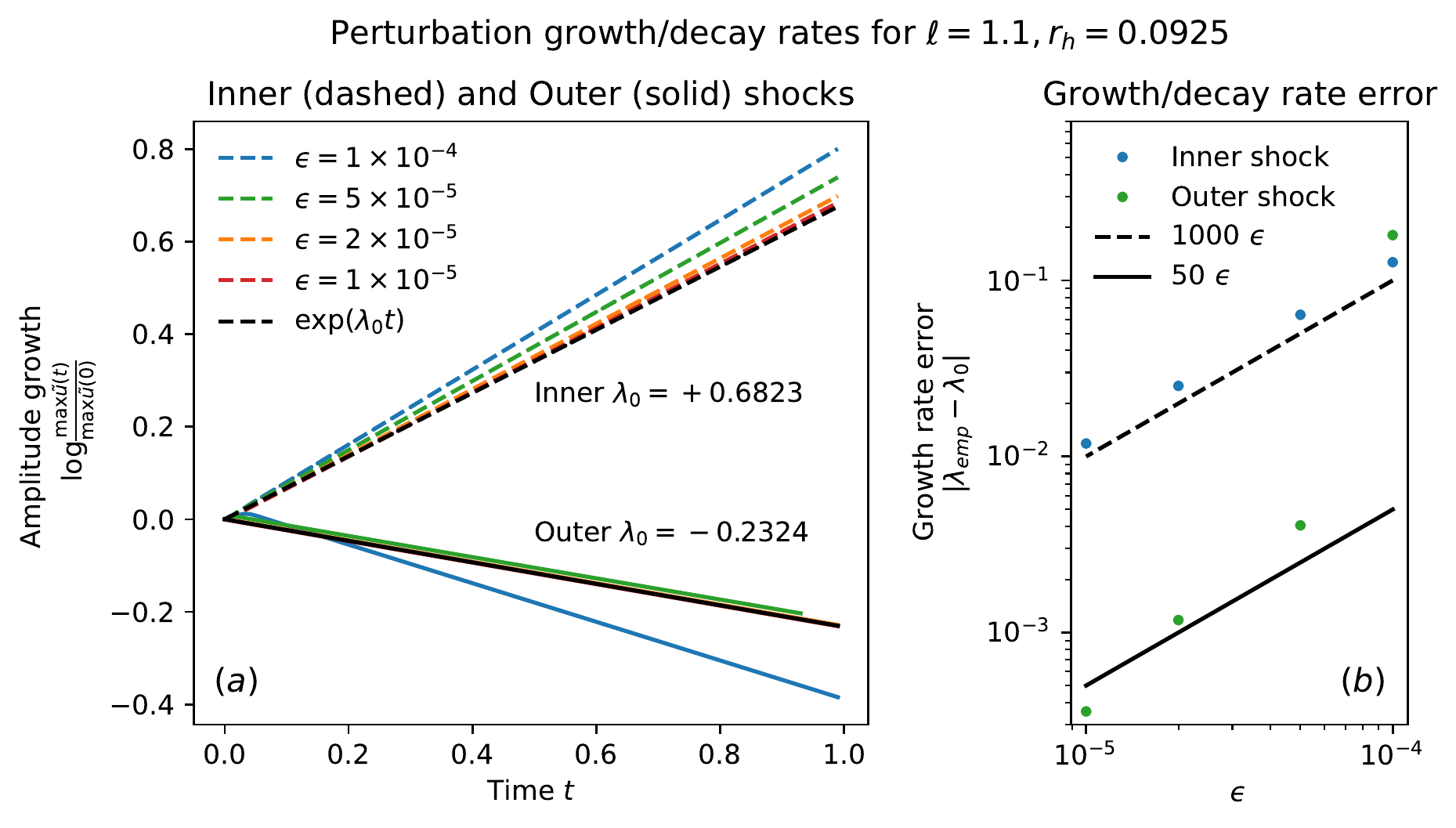}
    \caption{Empirical validation of asymptotic convergence to analytical growth rate $\lambda_0$ as $\eps \to 0$ at $\ell = 1.1, r_h = 0.0925$.
    In figure $(a)$ we plot the logarithm of the normalised peak magnitude of the velocity perturbation $\tilde u$ over time, for the inner (dashed) and outer (solid) shocks.
    We plot the analytical growth/decay rate calculations in black.
    In figure $(b)$ we quantify the error of the inner (blue) and outer (green) growth rates as $\eps \to 0$ on a log-log plot.
    We observe approximately $\O(\eps)$ convergence to analytical $\lambda_0$.
    }
    \label{fig:black-hole-shock-growth-rate-errors}
\end{figure}

We validate the asymptotic calculations by simulating the nonlinear perturbation equations \cref{eq:unsteady-nonlinear-delta,eq:unsteady-nonlinear-u,eq:unsteady-nonlinear-l} using Dedalus, 
where steady components ${}{u}, {}{\delta}, {}{\ell}$ are determined from the nonlinear steady problem (\cref{sec:steady-shock-numerics}).
We solve for the perturbations because this formulation is less stiff than solving for the full quantities, 
as nonconstant coefficients can be treated implicitly.
We solve for the evolution of inner and outer shocks at $\ell = 1.1$, $r_h = 0.0925$, for $\eps = \num{1e-4}, \num{5e-5}, \num{2e-5}, \num{1e-5}$.
We use $n = 256$ Chebyshev polynomials over the domain $(r_0,r_1) = (r_s + \eps (x_0 - 100), r_s + \eps (x_0+100))$, with $3/2$ dealiasing.
We apply zero Dirichlet boundary conditions on the perturbations at either endpoint, 
	\begin{align}
	\tilde u(t, r_0) = \tilde u(t, r_1) = \tilde \delta(t, r_1) = \tilde \ell(t, r_0) = \tilde \ell(t, r_1) = 0.
	\end{align}
The initial perturbations represent a small displacement of the shock
	\begin{align}
	\tilde u(0, r) &= \frac{\eps}{10} {} u'(r), &
	\tilde \delta(0, r) &= \frac{\eps}{10} {} \delta '(r), &
	\tilde \ell(0, r) &= \frac{\eps}{10} {} \ell'( r).
	\end{align}
We summarise the asymptotic convergence of the shock stability calculations as $\eps \to 0$ in \cref{fig:black-hole-shock-growth-rate-errors}.
\Cref{fig:black-hole-shock-growth-rate-errors} $(a)$ plots the normalised change in peak magnitude of $\tilde u$ for each simulation, with clear convergence to the analytic growth rates for the inner (dashed black) and outer (solid black) shocks.
\Cref{fig:black-hole-shock-growth-rate-errors} $(b)$ shows that this rate of convergence is linear in $\eps$.
We are thus confident in the validity of our asymptotic calculations.

\section{Discussion}	
\label{sec:discussion}
The Rankine-Hugoniot shock conditions neglect microscopic dissipation.
This omission has impeded previous linear stability analyses.
Earlier works determined bounds on growth/decay rates of shock instabilites \cite{NakayamaHydrodynamicInstabilityAccretion1992,NakayamaHydrodynamicInstabilityAccretion1993,NakayamaDynamicalInstabilityStanding1994},
corresponding to the range
	\begin{align}
    &\frac{1}{u^-_0(r_s) + u^+_0(r_s)} \br{\frac{1}{r} - \frac{2}{(r-r_h)^2} + \frac{\ell^2}{r^3}} \nonumber\\
    &\hspace{2cm} \leq \lambda_0 \leq
    \frac{1}{u^-_0(r_s) - 1} \br{\frac{1}{r} - \frac{2}{(r-r_h)^2} + \frac{\ell^2}{r^3}}.
	\end{align}
Specific growth rates $\lambda_0$ can only be resolved by accounting for microscopic dissipation.

The drawback of dissipation is that it increases the mathematical order of the problem.
But multiple scales matched asymptotics provide a straightforward procedure to solve singular perturbation problems,
giving an explicit formula for the growth/decay rates,
	\begin{align}
	\label{eq:growth-rate}
    \lambda_0 = \frac{1}{u^+_0(r_s) + u^-_0(r_s)}\br{\frac{1}{r_s}-\frac{2}{\left(r_s	-r_h\right){}^2} + \frac{\ell	^2}{r_s^3}}.
	\end{align}
This allows a description of shock stability throughout parameter space (\cref{fig:black-hole-accretion-regime-diagram,fig:black-hole-shock-stability-regimes}).

\begin{figure}[t]
	\centering
    \includegraphics[width=.7\linewidth]{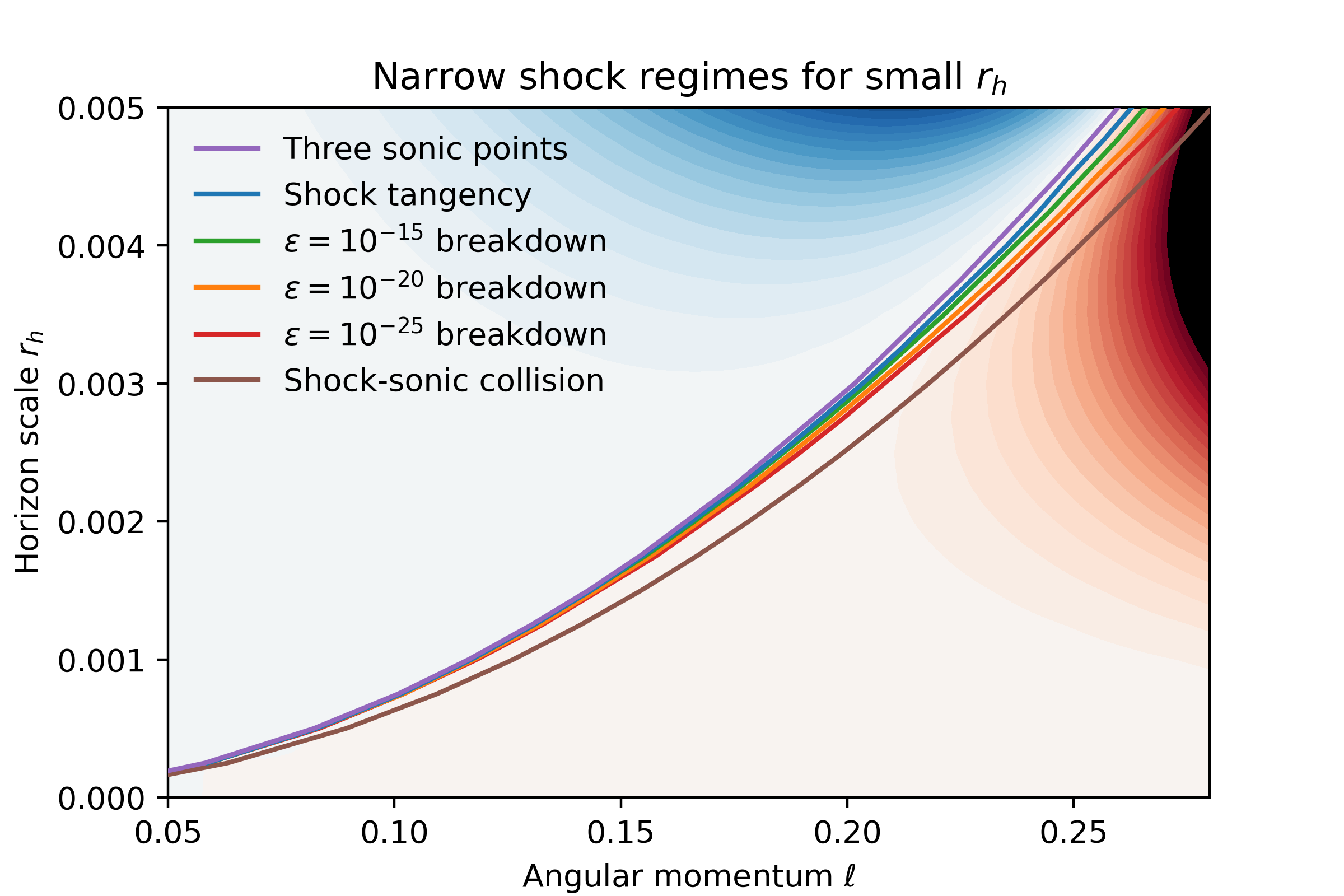}
    \caption{Regime diagram for realistic $r_h$. There are one/two sonic points in the blue/red regions respectively. 
    The boundary is plotted in purple.
    Shocks exists between the blue and brown lines.
    We also plot the level curves for which the maximum magnitude of $|u_1^-/u_0^-|$ equal to $10^{15}, 10^{20}, 10^{25}$ (green, orange, red respectively).
    The steady model breaks once $|u_1^-/u_0^-| > \eps^{-1}$.
    }
    \label{fig:black-hole-small-rh-asymptotic-breakdown-regimes}
\end{figure}

\textbf{Isothermal shock regimes are narrow} --- 
Realistic isothermal black hole shocks are restricted to a narrow sliver of parameter space (\cref{fig:black-hole-accretion-regime-diagram,fig:black-hole-small-rh-asymptotic-breakdown-regimes}).
Consider Cygnus X-1.
Assuming an ideal monatomic gas with sound speed and viscosity proportional to $T^{1/2}$, and a maximum possible temperature corresponding to the thermal emission cutoff at $\SI{100}{keV} \approx \SI{1e9}{K}$ \cite{GierlinskiSimultaneousXray7ray1997}, it is almost certain that $r_h \lesssim \num{5e-3}$ (and likely at least an order of magnitude smaller).
If we consider $r_h = \num{5e-3}$, then the total range where there are two possible shocks is $\ell \in (0.262, 0.281)$.
The range shrinks to $\ell \in (0.083, 0.089)$ for $r_h = \num{5e-4}$ (\cref{fig:black-hole-small-rh-asymptotic-breakdown-regimes}).
\begin{table}
\centering
  \begin{tabular}{cll}
    Symbol & Value & Dimensions\\
      \hline \\[-2.5ex]
    $G$ & 
    \SI{6.7e-11}{} &
    \si{m^3.kg^{-1}.s^{-2}} \\
    $M$ & 
    \SI{4.2e31}{} &
    \si{kg} \\
    $c$ & 
    \SI{3e8}{} &
    \si{m.s^{-1}} \\
    $T$ & 
    $10^9$ & % 100keV
    \si{K} \\
    $K$ & 
    \SI{3.7e5}{} &
    \si{m^2.s^{-2}} \\
    $\nu_2$ &
    $10^{-3}$&
    \si{m^2.s^{-1}} \\
    $r_h$ & 
    \num{5e-3}  &\\
    $\eps$ & 
    $10^{-20}$  &
    \end{tabular}
  \caption{Non-dimensional and dimensional parameters estimated for Cygnus X-1. $M$ is taken from \cite{Miller-JonesCygnusX1Contains2021}.}
\end{table}

The first order asymptotic problem further constrains this narrow regime.
While a kinematic viscosity of $\eps = 10^{-20}$ seems negligible, the smallness of $\eps$ is offset by the immensity of $|u^-_1/u^-_0|$.
Even infinitesimal angular momentum dissipation after the shock leads to enormous perturbations in the first order velocity $u^-_1$.
This highlights the importance of considering angular momentum perturbations in the asymptotic analysis even though the leading order angular momentum is constant.
Because $u^-_1$ has opposite sign from $u^-_0$,
once $|u_1^-/u_0^-| > \eps^{-1}$, the velocity $u^- = u_0^- + \eps u_1^-$ becomes zero, and the steady model breaks down.
Steady shocks at finite viscosity are only possible if $|u_1^-/u_0^-| < \eps^{-1}$.
\Cref{fig:black-hole-small-rh-asymptotic-breakdown-regimes} summarises shock regimes for $r_h < \num{5e-3}$.
While ideal shocks are possible between the blue and brown lines,
shocks at finite viscosity $\eps \approx 10^{-20}$ are only possible in a much narrower regime above the orange line.
Furthermore, this regime gets proportionately narrower as $r_h$ decreases.
\Cref{sec:smooth-stability} shows that the smooth transonic solution is always linearly stable.
This model therefore predicts that steady isothermal shocks are unlikely.
Additional physics would be necessary to find robust theoretical evidence of shocks in black hole accretion disks.

\section{Conclusions}
\label{sec:conclusion}
Previous models of isothermal shocks in black hole accretion disks predicted two possible steady shocks \cite{ChakrabartiStandingShocksIsothermal1989}, and found bounds on their stability \cite{NakayamaHydrodynamicInstabilityAccretion1992}.
We provide precise values for shock wave stability properties by considering microscopic dissipation.
Using multiple scales matched asymptotic expansions, we calculate the first explicit formula for growth/decay rates of isothermal shocks (\cref{eq:growth-rate}).
We support our asymptotic calculations using Mathematica, and provide numerical validation of these properties using Dedalus \cite{BurnsDedalusFlexibleFramework2020}.
We show that the inner shock is unstable, the outer shock is stable, and that growth/decay rates tend to zero as shocks approach the sonic points.
Using a simple energy argument, we find that the smooth transonic flow is also stable, suggesting two possible physical solutions.
However, our asymptotic analysis suggests that realistic black holes only support steady shock waves in extremely narrow parameter regimes, arguing against the existence of steady isothermal shocks for most black holes.
Better models of shocks in accretion disks require additional physics, such as non-isothermal gases, radiation, and magnetohydrodynamic dissipation. 
Fortunately, the techniques developed within this paper are general, and can in future help understand shocks in more complex astrophysical models.

\section*{Acknowledgements}
EH \& MW acknowledge support from ARC DP180103022 and DP200102130 grants.

%%%%%%%%%%%%%%%%%%%%%%%%%%%%%%%%%%%%%%%%%%%%%%%%%%
\section*{Data Availability}
All code (Mathematica notebook and python scripts) and data used in this investigation are available online at \href{https://github.com/ericwhester/isothermal-accretion-disk-shocks}{github.com/ericwhester/isothermal-accretion-disk-shocks}.

%%%%%%%%%%%%%%%%%%%% REFERENCES %%%%%%%%%%%%%%%%%%

\bibliographystyle{siam}
\bibliography{black-hole-references}

\appendix

\section{Steady viscous shock waves}
\label{sec:steady-asymptotics}
\subsection{Zeroth-order inner problem}
The leading order reduced problem cannot predict shocks.
To understand them we must rescale our problem around these shocks.
The leading order behaviour for the layer problem becomes
	\begin{align}
    \br{u_0 - \frac{1}{u_0}}\pd_x u_0 - 2 \br{\pd_x^2 u_0 - \frac{\pd_x u_0^2}{u_0}} &= u_0 \pd_x \br{u_0 + \frac{1}{u_0} - 2\frac{\pd_x u_0}{u_0}} = 0,
    \end{align}
which can be integrated once to find
	\begin{align}    
    u_0^2 -c u_0 + 1 - 2 \pd_x u_0 &= 0,
	\end{align}
where $c = u_0(\infty) + 1/u_0(\infty) = u^+_0(r_s) + u^-_0(r_s)$ by the asymptotic matching conditions, giving
	\begin{align}
    u_0 &= \frac{1}{2}\br{u_0^+ + u_0^- + (u_0^+ - u_0^-) \tanh\br{-\frac{1}{4}(u_0^+ - u_0^-)(x - x_0)}}.
	\end{align}
The angular momentum problem becomes
	\begin{align}
    \br{u_0 + \frac{\pd_x u_0}{u_0}} \pd_x \ell_0 - \pd_x^2 \ell_0 = 0,
	\end{align}
with the general solution
	\begin{align}
    \ell_0(x) = c_1 + c_2 \exp \br{\int^x u_0 \, dy}.
	\end{align}
The asymptotic matching conditions for $\ell_0$ require constant limiting behaviour as $x\to\pm\infty$, but the function behaves like $\exp({|u_0^+|| x|})$ as $x\to-\infty$.
The only way to satisfy the asymptotic matching boundary conditions is for $c_2 = 0$ and $c_1 = \ell^+_{0}(r_s) = \ell$. 
The specific angular momentum is therefore conserved across a shock.
We abbreviate the solutions as
	\begin{align}
    u_0(x) &= -\cosh a - \sinh a \tanh\br{\frac{\sinh (a)}{2} \, x}, &
    \ell_0(x) &= \ell,
	\end{align}
where we have defined the constant $a \equiv \log (- u^+_0(r_s))>0$.

\subsection{First-order outer problem}
The first order reduced problem is linear, and simplifies upon substitution of $\ell_0 = \ell$
	\begin{align}
	\pd_r\br{\br{u_0^\pm-\frac{1}{u_0^\pm}} u_{r,1}^\pm}
	-\frac{2 \ell }{r^3} \ell_1^\pm
	&=
	2 {u_0^\pm}''
	-\frac{2 \left({u_0^\pm}'\right){}^2}{u_0^\pm}
	-\frac{2 u_0^\pm}{r^2},
	\\
	{\ell_1^\pm}' 
	&=
	\frac{2\ell}{r u_0^\pm}\br{\frac{1}{r} + \frac{u_0^{\pm'}}{u_0^\pm}} = 2 \ell {\rho_0^\pm}'.
	\end{align}
The angular momentum equation can be integrated to find
    \begin{align*}
    \ell_1^\pm &= 2\ell \rho_0^\pm + c_1^\pm.
	\end{align*}
We then apply boundary conditions to complete the system.
The limiting boundary condition $\lim_{r\to\infty}\ell_1^+\to 0$ implies $c_1^+ = \lim_{r\to\infty}\rho_0^+ = \rho_\infty = \exp(e^*)$, the exponential of the specific energy at the outer sonic point.
At a sonic point $r_{*}$ the problem simplifies, and we can substitute the values of $u_0(r_*) = -1$ and its derivatives to determine a value for $u_1(r_*)$
	\begin{align}
    u_1(r_*) &= \left.\frac{1}{\pd_r u_0}\br{ \pd_r^2 u_0 + \pd_r u_0^2 + \frac{1}{r^2} + \frac{\ell}{r^3}\ell_1(r)}\right|_{r=r_*}.
	\end{align}

\subsection{First-order inner problem}
The inner system at first order can be written in terms of $u_1$ alone
	\begin{align}
	L[u_1] \equiv u_1'' - \frac{1}{2}\left(u_0-\frac{1}{u_0} + \frac{4 u_0'}{u_0}	\right)	u_1' +\cosh a \frac{u_0'}{u_0} u_1 = g(r_s),
	\end{align}
where 
    \begin{align}
    g(r_s) = -\frac{1}{2}\br{\frac{1}{r_s}-\frac{2}{\left(r_s	-r_h\right){}^2} + \frac{\ell	^2}{r_s^3}}.
    \end{align}
To proceed we analyse the linear operator $L$.

\textbf{First kernel component} $u_{1,1}$ --- 
The zeroth order layer problem was translation symmetric.
It follows that the derivative of the zeroth order solution is in the kernel of this operator,
	\begin{align}
	L[u_{1,1}] &= 0, \qquad
    u_{1,1} = u_0' = -\frac{\sinh^2 a}{2} \sech^2 \br{\frac{\sinh a}{2} x}.
	\end{align}

\textbf{Wronskian} $W$ ---
To find the second kernel component $u_{1,2}$ we calculate the Wronskian ${W = u_{1,1} u_{1,2}' - u_{1,1}' u_{1,2}}$.
If $L[y] = y'' - p(x) y' - q(x) y$, then $W' = p(x) W$.
Hence
	\begin{align}
	W & \propto \text{sech}^2\left(\frac{\sinh a x}{2}\right)   \left(-2 \cosh a - 2 \sinh a \tanh \left(\frac{\sinh a\,   x}{2}\right)\right)\nonumber\\
		&\propto 2 \operatorname{csch}^2 a \, u_0' u_0.
	\end{align}

\textbf{Second kernel component} $u_{1,2}$ --- 
It is then straightforward to solve the Wronskian for $u_{1,2}$
	\begin{align}
	u_{1,2} &\propto  u_0 + x u_0' + \sech(a) \nonumber\\
		&\propto -\sinh a \tanh   \left(\frac{1}{2} x \sinh   a\right)-\frac{1}{2} x   \sinh ^2 a   \sech^2\left(\frac{1}   {2} x \sinh   a\right)\nonumber\\
		 &\qquad -\cosh   a+\sech a.
	\end{align}
If we pick the preferred elements
	\begin{align}
    u_{1,1} &= u_0', &
    u_{1,2} &= u_0 + x u_0' + \sech a,
	\end{align}
then the Wronskian becomes
	\begin{align}
    W &=\sinh a \tanh a \, u_0' \, u_0.
	\end{align}

\textbf{Inhomogeneous solution} $u_{1,i}$ --- 
We then solve the particular solution $L[u_{1,i}] = 1$ using variation of parameters,
	\begin{align}
	u_{1,i} &= c_1(x) u_{1,1} + c_2(x) u_{1,2}, 
	\end{align}
where
    \begin{align}
	L[c_1 u_{1,1} + c_2 u_{1,2}] &= 1, &
	c_1' u_{1,1} + c_2' u_{1,2} &= 0.	
	\end{align}
Simplifying the constraints shows
	\begin{align}
    c_1' &= -\frac{u_{1,2}}{W}, &
    c_2' &= \frac{u_{1,1}}{W},
	\end{align}
which can be integrated to find
	\begin{align*}
	c_1(x) &= \frac{1}{\sinh a \tanh a}\left(2 x
		+\frac{1}{2} e^{-a}x^2\right. %\nonumber\\
		-2 x \log \left(e^{-x \sinh	a-2 a}+1\right)\nonumber\\
		&\hspace{2.3cm}-4	\sech a \log \left(\cosh	\left(\frac{1}{2} x \sinh a+a\right)\right)\nonumber\\
		&\hspace{2.3cm}+\operatorname{csch}^2	a \sech a \cosh (x \sinh	a)\nonumber\\
		&\hspace{2.3cm}+2 \operatorname{csch}a	\operatorname{Li}_2\left(-e^{-2 a-x \sinh	a}\right)\bigg).\\
    c_2(x) &=\frac{1}{\sinh a \tanh a}\br{ 2 \log \left(\cosh \left(\frac{1}{2}   x \sinh a + a\right)\right)- x   \cosh a}.
		\end{align*}
where $\operatorname{Li}_2$ is the polylogarithm of order 2.
The full solution $u_1$ combines the inhomogeneous solution $u_{1,i}$ and the two kernel components
	\begin{align}
    u_1 = g(r_s) u_{1,i} + b_1 u_{1,1} + b_2 u_{1,2}.
	\end{align}
To determine the coefficients $b_1,b_2$ of each kernel component requires further work.
The coefficient of $u_{1,2}$ is constrained by the limiting boundary conditions of the first order problem, which requires knowledge of both $\pd_r u^\pm_0(r_s)$ and $u^\pm_1(r_s)$.
As $u_{1,1}$ decays exponentially fast toward the boundaries, it cannot be determined by boundary conditions at any order. 
Instead it is determined through a solvability condition at subsequent order.
Such solvability conditions are most easily determined for self-adjoint operators.

\textbf{Self-adjoint} $L^\dag$ ---
We recast in self-adjoint form by reweighting $L$ with the Wronskian $W$,
	\begin{align}
	L^\dag [v] &= \br{\frac{v'}{u_0' u_0}}'+\frac{\cosh a}{u_0^2}v = \frac{1}{u_0' u_0}L[v],
	\end{align}
where $L^\dag$ is a Sturm-Liouville operator.
Hence the kernel and cokernel of $L^\dag$ are the same, simplifying solvability conditions.

\textbf{Zeroth order shift $x_0$} ---
Given the limiting behaviour of the solution we write
	\begin{align*}
	u_1 &= g(r_s) u_{1,i}(x_0) + b_1 u_{1,2}(x_0),\\
	u_{1,i} &= c_1 u_{1,1} + c_2 u_{1,2},\\ 
	u_{1,2} &= u_{1,2} = u_0 + x u_0' + \sech a.
	\end{align*}
The boundary conditions require
	\begin{align*}
	u_1(-X,x_0) &\sim u_1^-(r_s) - X \pd_r u_0^-(r_s), \nonumber\\
	u_1(X,x_0) &\sim u_1^+(r_s) + X \pd_r u_0^+(r_s).
	\end{align*}
The limiting linear behaviour of $u_{1,i}$, and the limiting constant behaviour of $u_{1,2}$ imply
	\begin{align*}
	u_1(-X,x_0) 
		&= g(r_s) u_{1,i}(-X, x_0) + b_2 u_{1,2}(-X, x_0),\nonumber\\
		&= g(r_s) (u_{1,i}(-X, 0) - x_0 u_{1,i}'(-X)) + b_2 u_{1,2}(-X,0),\nonumber\\
	u_1( X,x_0) 
		&= g(r_s) u_{1,i}(X, x_0) + b_2 u_{1,2}(X, x_0),\nonumber\\
		&= g(r_s) (u_{1,i}(X, 0) - x_0 u_{1,i}'(-X)) + b_2 u_{1,2}(X,0).
	\end{align*}
This reduces to a linear system for $x_0$ and $b_2$
	\begin{align*}
	&\begin{bmatrix}
	-g(r_s) u_{1,i}'(-X) & u_{1,2}(-X)\\
	-g(r_s) u_{1,i}'( X) & u_{1,2}(X)
	\end{bmatrix}
	\begin{bmatrix}
	x_0\\
	b_2
	\end{bmatrix}
	\nonumber\\
	&\hspace{2cm} =
	\begin{bmatrix}
	u_1^-(r_s) - X \pd_r u_0^-(r_s) - g(r_s) u_{1,i}(-X,0)\\
	u_1^+(r_s) + X \pd_r u_0^+(r_s) - g(r_s) u_{1,i}(X,0)
	\end{bmatrix}.
	\end{align*}

\textbf{Angular momentum equation} ---
The $\ell_1$ equation, after substituting for $\ell_{0}$, becomes
	\begin{align}
	\ell_1'' - \br{u_0 + \frac{u_0'}{u_0}}\ell_1' &= -\frac{2\ell}{r_h} \br{\frac{u_0'}{u_0}}.
	\end{align}
We then find
	\begin{align}
    \ell_1(r) 
    	&= \frac{2\ell}{r_h}\br{e^{\int^r u_0 ds} \int^r e^{-\int^s u_0 dt} ds} + c_1,\nonumber\\
    	&= \frac{\ell}{r_h}\br{\sech a + \cosh(a - x\sinh a) } \sech\br{\frac{\sinh a}{2}x }^2 + c_1,
	\end{align}
which behaves like a tanh profile plus a constant determined by asymptotic matching conditions.
This jump is precisely that predicted by the jump in density for the first order outer equation, meaning that $\ell^\pm_1 = 2\ell (\rho^\pm - \rho_\infty)$ for both inner and outer solutions.

\section{Smooth supersonic stability analysis}
\label{sec:smooth-supersonic}

We show that smooth inviscid supersonic flow within the sonic point is stable by showing that all normal modes with homogeneous Dirichlet boundary conditions at the sonic point have negative growth rates.
Sufficiently small viscous perturbations will not change this spectral property, i.e. these growth rates stay negative.

We have the following asymptotic behaviour for $r, x$, and  ${}{u}$ as $r\to r_*$ from below,
    \begin{align}
    {}{u} &\approx -1 + (r - r_*)\pd_r {}{u},\nonumber\\
    \frac{dx}{dr} &= \frac{-{}{u}}{{}{u}^2 - 1} \approx \frac{1}{2\pd_r {}{u} (r_* - r)},\nonumber\\
    x(r) &\approx \frac{-\log(r_* - r)}{2\pd_r {}{u}},\nonumber\\
    r_* - r &\approx e^{-2\pd_r {}{u} x},\nonumber\\
    \tilde{u} &\approx \frac{e^{2\pd_r {}{u} x}}{2\pd_r {}{u}}\pd_x \psi.
    \end{align}
This leads to the following asymptotic behaviour for the $w$ equation
    \begin{align}
    w'' &\approx \lambda^2 (1 - 2 \pd_r {}{u} e^{-2 \pd_r {}{u} x})w.
    \end{align}
We can determine limiting behaviour for the solution $w$ (and the physical perturbation velocity $\tilde{u}$) using WKB analysis.
We assume the following asymptotic behaviour for $w$,
    \begin{align}
    w &= A e^{-|\lambda|x - f}, \quad f > 0, \quad f\to 0 \text{ as } x \to \infty,
    \end{align}
which when substituted into the $w$ equation implies
    \begin{align}
    w'' &= A(-f'' + f'^2 + 2 |\lambda| f' + |\lambda|^2)w,\nonumber\\
        &\approx A(-f'' + 2|\lambda| f' + |\lambda|^2)w.
    \end{align}
We can then solve for the decaying term $f$ to find
    \begin{align*}
    -f'' + 2 |\lambda| f' &= - 2\pd_r {}{u} |\lambda|^2 e^{-2\pd_r {}{u} x},\\
    f &= \frac{|\lambda|^2}{2(|\lambda| + \pd_r {}{u})}e^{-2\pd_r {}{u} x},\\
    w &\approx e^{-|\lambda| x} \exp\br{-\frac{\lambda^2}{2(|\lambda|+\pd_r {}{u})}e^{-2 \pd_r {}{u} x}}.
    \end{align*}
This then implies that the physical perturbation velocity $\tilde{u}$ has the following limiting behaviour as $r\to r_*$
    \begin{align*}
    \psi &\approx  \exp\br{(\lambda - |\lambda|)x-\frac{\lambda^2}{2(|\lambda|+\pd_r {}{u})}e^{-2 \pd_r {}{u} x}},\\
    \tilde{u} &\approx \frac{\lambda^2}{\lambda + \pd_r {}{u}}\exp\br{-\frac{\lambda^2}{2(|\lambda|+\pd_r {}{u})}e^{-2 \pd_r {}{u} x}},\\
    &\to \frac{\lambda^2}{\lambda + \pd_r {}{u}} > 0 \quad \text{ as } \quad r \to r_* \quad \text{ if } \quad \lambda > 0.
     \end{align*}
Modes with non-negative growth rate therefore have non-zero boundary conditions at the sonic point, and so all normal modes with zero boundary conditions at the sonic point must have negative growth rates.
Smooth transonic accretion flows are therefore linearly stable.

\section{Shock stability analysis}
\label{sec:stability-asymptotics}
We determine which possible solutions (inner shock, outer shock, or smooth transonic flow) are physical by determining their linear stability.

\subsection{Shock stability analysis}
To anlayse the shock stability problem we must zoom in to the shock length scale, following the same asymptotic procedure as for the steady problem.

\textbf{Zeroth order equations} ---
The leading order problem in $\eps$ reduces to 
	\begin{align}
	{}{u}_0 \pd_x \tilde{\delta}_0
	-\frac{\tilde{u}_0 {}{u}_0'}{{}{u}_0}
	+\pd_x\tilde{u }_0 &= 0\nonumber\\
	{}{u}_0 \pd_x\tilde{u}_0
	+{}{u}_0' \tilde{u}_0
	+\pd_x\tilde{\delta }_0
	+\frac{2 {}{u}_0' \pd_x\tilde{u}_0}{{}{u}_0}
	-2 {}{u}_0' \pd_x\tilde{\delta}_0
	-2 \pd_x^2 \tilde{u}_0 &= 0\nonumber\\
	\br{{}{u}_0 + \frac{{}{u}_0'}{{}{u}_0}}\pd_x \tilde{\ell}_0
	-\pd_x^2 \tilde{\ell}_0 &= 0
	\end{align}
	
\textbf{First order equations} ---
At the next order the equations satisfy
	\begin{align}
	&{}{u}_0 \pd_x\tilde{\delta}_1
	-\frac{\tilde{u}_1 {}{u}_0'}{{}{u}_0}
	+\pd_x\tilde{u}_1
	= -\br{\pd_t\tilde{\delta }_0
	+{}{u}_1 \pd_x \tilde{\delta}_0
	-\br{\frac{{}{u}_1}{{}{u}_0}}'\tilde{u}_0},\nonumber\\
	&{}{u}_0 \pd_x\tilde{u}_1
	+\tilde{u}_1 {}{u}_0'
	+\pd_x\tilde{\delta}_1
	+\frac{2 {}{u}_0' \pd_x\tilde{u}_1}{{}{u}_0}
	-2 {}{u}_0' \pd_x\tilde{\delta }_1
	-2\pd_x^2 \tilde{u}_1 = \nonumber\\
	&\hspace{.5cm}-\bigg(\pd_t\tilde{u}_0
	+{}{u}_1'\tilde{u}_0 
	+{}{u}_1 \pd_x\tilde{u}_0  
	-2 {}{u}_1' \pd_x \tilde{\delta }_0
	+2\br{\frac{{}{u}_1}{{}{u}_0}}' \pd_x \tilde{u}_0\bigg),\nonumber\\
	&\br{{}{u}_0 + \frac{{}{u}_0'}{{}{u}_0}} \pd_x\tilde{\ell}_1 - \pd_x^2\tilde{\ell}_1 
	= - \br{\frac{2\ell}{r_h}\br{\frac{{}{u}_0'}{{}{u}_0}}' + \br{{}{u}_0 - \frac{{}{u}_0'}{{}{u}_0}}' {}{\ell}_1' },
	\end{align}
The left-hand side is a consistent linear operator applied to the highest order terms.
The right-hand side represents source terms from higher order problems.
These source terms are constrained by solvability conditions on the left-hand linear operators.

\subsection{Zeroth order shock instability problem}
If we put in the ansatz
 	\begin{align}
	\tilde{u}_0(t,x) &= e^{\lambda_0 t} {}{u}_0'(x), &
	\tilde{\delta}_0(t,x) &= e^{\lambda_0 t} {}{\delta}_0'(x) &
	\tilde{\ell}_0(t,x) &= 0,
	\end{align}
then the zeroth order equations are satisfied.
The spatial profile thus corresponds to translation of the shock location at rate $\lambda_0$.

\subsection{First order shock instability problem}
Substituting these solutions into the next order gives the equations
	\begin{align}
	&{}{u}_0 \tilde{\delta	}_1'
	-\frac{{}{u}_0'}{{}{u}_0} \tilde{u}_1
	+\tilde{u}_1' =
		\lambda_0 \frac{{}{u}_0'}{{}{u}_0}
		+ {}{u}_0 \br{ \frac{{}{u}_0'}{{}{u}_0} \frac{u_1}{u_0}}',\nonumber\\
	& {}{u}_0   \tilde{u}_1'
	+\tilde{u}_1 {}{u}_0'
	+\tilde{\delta }_1'
	+2\br{\frac{{}{u}_0'   \tilde{u}_1'}{{}{u}_0}
	-{}{u}_0' \tilde{\delta   }_1'
	- \tilde{u}_1''} =\nonumber\\
	&\hspace{1cm}
		-\lambda_0 {}{u}_0'
		-\br{{}{u}_0' {}{u}_1}'
		-2 \br{\br{\frac{{}{u}_0'}{{}{u}_0}}' {}{u}_1' + \br{\frac{{}{u}_1}{{}{u}_0}}' {}{u}_0''}.
	\end{align}
The zeroth order angular momentum perturbation drops out of the analysis, and so does not contribute further to the stability calculation.

\textbf{Rearranging in terms of $u$} --- 
We can rearrange this problem to derive the same linear operator as the steady problem.
Substituting for $\tilde{\delta}_1$ and simplifying derivatives of ${}{u}_0$, we find
	\begin{align}
	&\left(\pd_x^2
	-\frac{1}{2}\br{{}{u}_0 - \frac{1}{{}{u}_0} + \frac{4{}{u}_0'}{{}{u}_0}}\pd_x
	+\cosh a \frac{{}{u}_0'}{{}{u}_0}
\right)\left[\tilde{u}_1+\lambda_0 \right]= \nonumber\\
    &\hspace{2cm} \frac{{}{u}_0'}{2{}{u}_0^2 } \left( \cosh a
	\left(1-{}{u}_0^2\right) {}{u}_1
	+\left(3 {}{u}_0^2-1\right)
	{}{u}_1'\right).
	\end{align}
	
\textbf{Self-adjoint form} --- 
To clarify solvability conditions (where we must integrate against the cokernel), we rewrite the problem in self-adjoint form
	\begin{align}
	L^\dag [\tilde{u}_1 + \lambda_0] &= \br{\frac{(\tilde{u}_1 + \lambda_0)'}{{}{u}_0' {}{u}_0}}'
	+\frac{\cosh a}{{}{u}_0^2}(\tilde{u}_1 + \lambda_0) \nonumber\\
	&=\frac{1}{2{}{u}_0^3} \br{\cosh a	\left(1-{}{u}_0^2\right) {}{u}_1
		+\left(3 {}{u}_0^2-1\right){}{u}_1'}.
	\end{align}

\textbf{Solving for $\lambda_0$} ---
We can expand the problem as
	\begin{align*}
	L^\dag[\tilde{u}_1] =-\cosh a \frac{\lambda_0}{{}{u}_0^2} + \frac{1}{2 {}{u}_0^3} \left(\cosh a	\left(1-{}{u}_0^2\right) {}{u}_1
		+\left(3 {}{u}_0^2-1\right){}{u}_1'\right).
	\end{align*}
Solvability requires that the right hand side is orthogonal to the cokernel (equal to the kernel for a self-adjoint operator)
which gives an explicit formula for $\lambda_0$
	\begin{align*}
    \lambda_0 &= \frac{\int_{-\infty}^\infty \frac{{}{u}_0'}{2{}{u}_0^3}\left(\cosh a	\left(1-{}{u}_0^2\right) {}{u}_1
		+\left(3 {}{u}_0^2-1\right){}{u}_1'\right)\,dx}{\cosh a\int_{-\infty}^\infty \frac{u_0'}{u_0^2}\, dx},
	\end{align*}
which simplifies to the concise form
	\begin{align}
    \lambda_0 = -g(r_s) \sech a = \frac{\frac{1}{r_s}-\frac{2}{\left(r_s	-r_h\right){}^2} + \frac{\ell^2}{r_s^3}}{u^+_0(r_s) + u^-_0(r_s)}.
	\end{align}
\end{document}